\documentclass{jfm}
\usepackage{graphicx}
\usepackage{epstopdf, epsfig}
\usepackage{amsmath}
\usepackage{amsfonts}
\usepackage{amssymb}
\usepackage{mathrsfs}
\usepackage{graphicx}

\usepackage{hyperref}
\usepackage{natbib}
\usepackage{dsfont}

\usepackage{color}
\usepackage{framed}
\usepackage{datetime}

\renewcommand{\vec}{\boldsymbol}
\renewcommand{\hat}{\boldsymbol}

\newcommand{\tens}[1]{\mathsfbi{#1}}
\newcommand{\indforce}{\vec{F}^\mathrm{ind}_\omega}
\newcommand{\extforce}{\vec{F}^\mathrm{ext}_\omega}
\newcommand{\dragforce}{\vec{F}^\mathrm{drag}_\omega}
\newcommand{\dragforcesubscript}{\vec{F}_\mathrm{drag}}

\newcommand{\constp}{p}
\newcommand{\constq}{q}
\newcommand{\admittance}{\tens{Y}_\omega}
\newcommand{\dragtensornoarg}{\tens{\pmb{\gamma}}}
\newcommand{\dragtensor}{\dragtensornoarg(\omega)}
\newcommand{\imagunit}{\mathrm{i\:\!}}
\newcommand{\thepi}{\upi\:\!}
\newcommand{\eulere}{\mathrm{e}}
\newcommand{\hide}[1]{}
\newcommand{\der}[1]{\:\frac{\mathrm{d}}{\mathrm{d}{#1}}\!}

\newcommand{\longshort}[2]{{#1}} 
\newcommand{\suppress}[1]{}

\shorttitle{Unsteady Stokes flow near boundaries}
\shortauthor{A. Simha, J. Mo and P. J. Morrison}

\title{Unsteady Stokes flow near boundaries: the point-particle approximation and the method of reflections}

\author{A. Simha\aff{1}
  \corresp{\email{akarsh@utexas.edu}},
  J. Mo\aff{2}
 \and P. J. Morrison\aff{1}}

\affiliation{\aff{1}Department of Physics and Institute for Fusion Studies, University of Texas at Austin, Austin, TX 78712, USA
\aff{2}Department of Physics and Center for Nonlinear Dynamics, University of Texas at Austin, Austin, TX 78712, USA}

\begin{document}

\maketitle

\begin{abstract}
  Problems of particle dynamics involving unsteady Stokes flows in confined geometries are typically harder to solve than their steady counterparts. Approximation techniques
  are often the only resort.~\citet[see e.g.][]{felderhof2005JPhysChemB,felderhof2009cylinder} has developed a point-particle approximation framework to solve such problems,
  especially in the context of Brownian motion. Despite excellent agreement with past experiments, this framework has an inconsistency which we address in this work. Upon
  implementing our modifications, the framework passes consistency checks that it previously failed. Further, it is not obvious that such an approximation should work for
  short time-scale motion. We investigate its validity by deriving it from a general formalism based on integral equations through a series of systematic approximations. We
  also compare results from the point-particle framework against a calculation performed using the method of reflections, for the specific case of a sphere near a full-slip
  plane boundary. We find from our analysis that the reasons for the success of the point-particle approximation are subtle and have to do with the nature of the unsteady
  Oseen tensor. Finally, we provide numerical predictions for Brownian motion near a full-slip and a no-slip plane wall based on the point-particle approximation as used by
  Felderhof, our modified point-particle approximation, and the method of reflections. We show that our modifications to Felderhof's framework would become significant for
  systems of metallic nanoparticles in liquids.
\end{abstract}


\tableofcontents

\section{Introduction}

The study of motion of particles in fluids has wide-ranging applications. Of interest here are problems that involve calculations of the resistance encountered by a rigid body
translating in a viscous incompressible fluid. If the motion of the body is sufficiently slow, it is often possible to approximate the flow of the fluid by steady Stokes flow.

\longshort{The problem of determination of the drag on a sphere in a fluid in the presence of other boundaries has been long studied in the context of steady Stokes
flow. \citep[For a selection of such results, see][]{lorentz1907stokeslet,faxen1921thesis,happel1965,alam1980,maul1996image}}{}

In the recent years, a new regime of viscous flow has gained substantial interest, wherein the Reynolds numbers are small, but the timescales of interest are comparable to or
shorter than the timescale of vorticity diffusion over the boundary. This is the regime of unsteady Stokes flow ~\citep[see e.g.][\S 1.1]{pozrikidis1992}. One application of
this regime is in the study of short time-scale Brownian motion, the exploration of which opens doors to the experimental study of statistical mechanics~\citep[see
e.g.][]{mo2015opex,kheifets2014science,franosch2011resonances},\longshort{}{ and} aids in the calibration of precision instrumentation such as optical
tweezers~\citep{bergsorensen2004powerspectrum,grimm2012PhysRevE.86.021912}\longshort{, and may provide a tool to measure the viscoelastic properties of complex fluids~\citep{felderhof2009viscoelasticity} and to probe boundary conditions on surfaces~\citep{lauga2005wettability,mo2017wettability}. Other applications include atomic force microscopy and microelectromechanical
systems~\citep{clarke2006microcantilever}}{}.

As the system can be approximated by linear equations in this regime, it is typical to study the problem of a particle performing small oscillations about a point. Despite the
linearity, however, these equations can be significantly harder to solve than the corresponding steady Stokes problems, particularly in situations with reduced symmetry. For
example, while the problem of a sphere translating near a plane wall may be solved by means of separable eigenfunction expansions in the case of steady Stokes
flow~\citep{oneill1964bispherical}, this is not true with unsteady Stokes flow\longshort{: the choice of coordinates that is apt for the symmetry of the problem is the bi-spherical
coordinate system, and the Helmholtz equation obtained by considering harmonic oscillations is not separable in this coordinate system, although the Laplace equation
is~\citep{morse1953methods}}{}. Thus, approximation techniques are inevitable.

~\citet{felderhof2005JPhysChemB,felderhof2006twowalls,felderhof2009cylinder,felderhof2012PhysRevE} has applied a point-particle approximation to determine the dynamics of a
sufficiently small spherical particle performing small oscillations in a number of confined geometries\footnote{Although Felderhof's work includes generalisations to
  compressible fluids, we restrict ourselves to incompressible fluids in this analysis}. In essence, his method involves approximating the spherical particle by a point
force for purposes of calculating the correction to the flow induced by the confining boundary. This results in a significant simplification of the original problem to what is, in
essence, a Green's function problem. However, it appears that Felderhof's result for a sphere near a plane wall does not reproduce the effective mass obtained from potential flow
calculations (see section~\ref{sec:potential_flow} for details). It also leads to a drag coefficient that depends on the density of the particle, which is inconsistent with
the fact that one may calculate the drag coefficient without any reference to the particle's density (\S\ref{sec:basic_setup}). Moreover, it is not obvious that the point-particle approximation
generalises to the case of unsteady Stokes flow. This is because of the existence of an additional length scale in the unsteady Stokes problem, namely the frequency-dependent
skin-depth of the vorticity shed by the particle. In an analysis of Brownian motion, there are fluctuations of all frequency scales, and therefore, this skin-depth cannot
always be assumed to be much larger than the particle size.

The first issue is related to determining the strength of the point force that reproduces the flow field of the sphere in the far field. In the case of steady Stokes flow, as
described by \citet[][\S 7]{lorentz1907stokeslet}, this is simply equal to $6\thepi \eta \vec{v}_s$, where $\vec{v}_s$ is the velocity with which the sphere translates. In the
unsteady case, Felderhof uses the external force $\extforce$ acting on the sphere as the point-force acting on the fluid. However, this produces a result that does not agree
in the far field, and as we stated earlier, results in a spurious dependence of the drag coefficient on the density of the particle. In this paper, we show that the point
force that reproduces the flow from a sphere in the far field is the induced force $\indforce$ described by \citet{mazur1974faxen}. Making this change in
Felderhof's theory results in correct values for the effective mass, and removes the spurious dependence of the drag coefficient on particle density.

As for the second issue of the existence of two length scales, we show by a systematic analysis of the approximation that there is a non-trivial reason why the approximation works in practice, as has been seen through its agreement with experiments~\citep{jeney2008boundary,mo2015pre}. To further enhance understanding of the approximation, we consider the simple case of a no-slip sphere of radius $a$ located at a distance $h$ from a
full-slip plane wall, and compare the results with an alternate calculation performed using the method of reflections. This alternate calculation results in a
drag coefficient that differs in the factors multiplying an exponential term. However, in the regime where this exponential term is important, the particle size is
indeed small in comparison with both length scales, viz. the distance from the wall and the skin depth of vorticity, whereby both methods produce similar results to leading order at all frequencies.

This paper is organised as follows. In section~\ref{sec:point_particle_approximation}, we review some well-known results and present our modifications to Felderhof's
point-particle approximation. In section~\ref{sec:validity}, we set up a general formalism from which we recover our modified version of Felderhof's framework through a series of systematic approximations. In section~\ref{sec:reflection_method}, we present the alternate calculation using the method of reflections for the simple case of a no-slip
sphere near a full-slip plane wall.  In section~\ref{sec:comparison_of_results}, we compare the results from the two methods, first by examining various limits, and then by
numerical evaluation. Thereafter, in section~\ref{sec:brownian_motion}, we apply the results to the hydrodynamic theory of Brownian motion. We conclude with a discussion in
section~\ref{sec:discussion}.

\section{The point-particle approximation}
\label{sec:point_particle_approximation}


\subsection{Computation of unsteady drag coefficients}
\label{sec:basic_setup}
We consider here the problem of a small rigid body $S$ of generic shape performing small translational oscillations in an arbitrary direction at arbitrary\footnote{It is assumed however that the frequency is not high enough that the compressibility of the fluid becomes important} frequency $\omega$ in an incompressible fluid of
dynamic viscosity $\eta$ and density $\rho_f$\footnote{In these problems, it is assumed for simplicity that the boundary of the particle itself does not change position, but
  the velocity boundary condition on that boundary changes. This results in a linear problem, and one would expect it to be good so long as the amplitude of oscillations is
  small and gets smaller as the frequency grows~\citep[see e.g.][]{zwanzig1970hydrodynamicVACF,mazur1974faxen}}. The fluid is bounded by various additional stationary surfaces (walls)
$W_i$ (which could have arbitrary shapes). The intention at a later stage will be to specialise $S$ to a sphere, and then consider a single plane wall $W$.

In many practical situations, one is interested in determining the net force (which we shall colloquially refer to as the ``drag force'') exerted by the fluid on the body. In order to do so, we wish to solve the unsteady incompressible Stokes equations,
\begin{equation}
\label{eq:UnsteadyStokes}
\begin{aligned}
\rho_f \frac{\partial \vec{v}}{\partial t} &= - \bnabla P + \eta \Delta \vec{v}, \\
\bnabla \bcdot \vec{v} &= 0,
\end{aligned}
\end{equation}
subject to some combination of no-slip or full-slip boundary conditions\footnote{We restrict ourselves to these special
  cases in this work.} on $\partial S$ and $W_i$. Here $\vec{v}(\vec{r},t)$ is the fluid velocity field, and $P(\vec{r}, t)$ is the pressure field.

We may Fourier transform the equations~\eqref{eq:UnsteadyStokes} in time to obtain
\begin{equation}
\label{eq:UnsteadyStokesFourier}
\begin{aligned}
 \Delta \vec{v}_\omega-\alpha^2 \vec{v}_\omega &= \frac{\bnabla P_\omega}{\eta}, \\
\bnabla \bcdot \vec{v}_\omega &= 0,
\end{aligned}
\end{equation}
where we have introduced the notation
\begin{equation}
\alpha := \sqrt{\frac{-\imagunit \omega \rho_f}{\eta}}\qquad\Real[\alpha] > 0,
\end{equation}
for the complex inverse skin-depth of vorticity\footnote{That this is an interpretation for $\alpha$ can be seen by taking the curl of the first equation
  in~\eqref{eq:UnsteadyStokesFourier}}, and $\vec{v}_\omega(\vec{r})$ and $P_\omega(\vec{r})$ are the Fourier transforms of $\vec{v}(\vec{r}, t)$ and $P(\vec{r}, t)$ respectively.\footnote{We use the convention $f_\omega = \int_{-\infty}^\infty \mathrm{d}t \; f(t) \; \eulere^{\imagunit \omega t}$ for Fourier transforms throughout this work.}

Once the solutions for $\vec{v}_\omega$ and $P_\omega$ have been computed, one may compute the drag force on the body as
\begin{equation}
\label{eq:DragForceSurfaceIntegral}
\dragforce = \oint_{\partial S} \mathrm{d}^2x\;\,\pmb{\sigma} \bcdot \hat{n},
\end{equation}
where $\pmb{\sigma}$ is the stress tensor having components $\sigma_{ij}(\vec{r}; \omega) = P_\omega \delta_{ij} + \eta(\partial_i v_{\omega j} + \partial_j v_{\omega i})$ and $\hat{n}$ is the outward unit normal to the
surface $\partial S$. Since the system is linear in the low Reynolds number regime, the drag force $\dragforce$ is a linear response to the velocity $\vec{u}_\omega$ of
the body, whereby it should be possible to write
\begin{equation}
\label{eq:DragCoefficientTensorDefinition}
\dragforce = - \pmb{\gamma}(\vec{r}_0; \omega) \bcdot \vec{u}_\omega,
\end{equation} where
$\pmb{\gamma}(\vec{r}_0; \omega)$ is a tensor of drag coefficients. Here, we have explicitly indicated that the drag coefficients are dependent on the position
$\vec{r}_0$ of the body, although we will drop this in the future to simplify notation.

We remark that the setup of this problem to compute the drag coefficient made no reference to the density of the body itself, and the effects of the body on the fluid were
captured through the boundary conditions at $\partial S$.

In general, analytically solving these equations in situations where the configuration of $S$ and $W_i$ does not possess sufficient symmetry poses difficulties as separable
eigenfunction expansions may not exist. \longshort{As mentioned earlier, e}{E}ven for the simple case of a sphere for $S$ and a single plane wall $W$, the Helmholtz equation (with
complex wavenumber) in~\eqref{eq:UnsteadyStokesFourier} is not separable in a coordinate system that is suitable for the symmetry of the boundaries. Thus, it is natural to
consider approximation techniques. The point-particle approximation~\citep{felderhof2005JPhysChemB}, matched asymptotic expansions~\citep{oneill1967matchedasymptotic}, and the
method of reflections~\citep[see e.g.][]{happel1965} are some approximation techniques to resort to.

\subsection{An overview of the point-particle framework of Felderhof}
\label{sec:felderhof_framework}
In this subsection, we review Felderhof's framework for computing particle dynamics using the point-particle approximation in general terms.  Felderhof has applied the
point-particle approximation to a number of situations~\citep[see e.g.][]{felderhof2005JPhysChemB,felderhof2006twowalls,felderhof2009cylinder,felderhof2012PhysRevE},
especially in the context of the hydrodynamic theory of Brownian motion. In this approximation, the body $S$ is replaced by a point force. This is in the spirit of a multipole
expansion~\citep[see e.g.][]{kim2013microhydrodynamics}, the idea being that in the far-field, the stokeslet part of the expansion dominates. Thus, for purposes of calculating
the effects of the walls $W_i$, it suffices to truncate the multipole expansion at the stokeslet level. Linearity
allows us to superpose the effects of the wall and the effects local to the body, a step that will later be effected using a generalised Fax\'{e}n theorem.

We begin by computing the vector-valued Green's function for the pressure field $\vec{P}$ (with components $P_j$) and tensor-valued Green's function for the velocity field
$\tens{G}$ (with components $G_{ij}$), arising from a general point force of unit strength at a generic location $\vec{r}'$. The Green's functions satisfy the equations
\begin{eqnarray}
\label{eq:StokesletProblemMomentum}
\Delta G_{ij}(\vec{r}|\vec{r}';\omega) -\alpha^2 G_{ij}(\vec{r}|\vec{r}';\omega) - \frac{1}{\eta}\partial_i P_j(\vec{r}|\vec{r}';\omega) &=& \delta_{ij} \delta(\vec{r}-\vec{r}'), \\
\label{eq:StokesletProblemConstraint}
\partial_i G_{ij}(\vec{r}|\vec{r}';\omega) &=& 0,
\end{eqnarray}
and also obey the required boundary conditions on the walls $W_i$. In principle, they may be computed by using the incompressibility condition in the first equation to
get the Poisson equation for the pressure $P_j$,
\begin{equation}
-\frac{1}{\eta} \Delta P_j(\vec{r}|\vec{r}';\omega) = \partial_j \delta(\vec{r}-\vec{r}'),
\end{equation}
and then substituting the solution of the above as a source into equation~\eqref{eq:StokesletProblemMomentum}. The resulting Helmholtz equations with complex wavenumber are then solved to determine
$G_{ij}$. In practice, the equations are generally solved using eigenfunction expansions and then applying boundary conditions to determine the coefficients~\citep{jones2004,felderhof2005JPhysChemB}.

The effect of the boundary conditions on the surface of the body $\partial S$ could in general be modelled by a force distribution (see section~\ref{sec:faxen_theorem}), which
could then be integrated against the above Green's function to obtain the velocity field. However, this is a non-trivial task in the complicated geometries of interest.  In
the point-particle approximation, the effect of the body $S$ is instead modelled by a single point force $\indforce$ at the location of the body\footnote{The problem of
  choosing this location is akin to finding a good choice for the origin in any multipole expansion} $\vec{r}_0$, which reproduces the flow from the actual body at
sufficiently large distances from the body. The change in the flow caused by the presence of the walls, may then be written as
\begin{equation}
  \label{eq:ReflectedFlowField}
  \vec{v}_W(\vec{r}|\vec{r}_0;\omega) = \left [ \tens{G}(\vec{r}|\vec{r}_0;\omega) - \tens{G}^0(\vec{r} - \vec{r}_0;\omega) \right ] \bcdot \indforce,
\end{equation}
where $\tens{G}^0$ is the free-space velocity Green's function ~\citep[i.e. the unsteady Oseen tensor; see e.g.][\S 6.2]{kim2013microhydrodynamics}. One may obtain $\tens{G}^0$ by the same method described to compute $G_{ij}$ except with the boundary condition being that the flow decay at infinity. The result may be written as~\citep{mazur1974faxen,felderhof2012PhysRevE}
\begin{equation}
\label{eq:UnsteadyOseenTensor}
\tens{G}^0(\vec{q}; \omega) = -\frac{1}{\eta} \Big ( G(\vec{q}; \omega) \mathds{1} + \alpha^{-2} \bnabla\bnabla \left [ G(\vec{q}; 0) - G(\vec{q}; \omega) \right ] \Big )
\end{equation}
Here, $G({\vec{q}}; \omega) = -\frac{\eulere^{-\alpha |{\vec{q}}|}}{4 \thepi |{\vec{q}}|}$ is the fundamental solution of the Helmholtz equation with complex $\alpha$ and $G({\vec{q}}; 0)$ is that of the Laplace equation.

We are yet to specify what
$\indforce$ must be to reproduce the flow generated by the body sufficiently far from it, and we shall do so in section~\ref{sec:induced_force}. Once the effect of the wall $\vec{v}_W$ is known, a generalised Fax\'{e}n theorem (section~\ref{sec:faxen_theorem}) may be used to compute the drag coefficient.

When using the generalised Fax\'{e}n theorem in the point-particle approximation, it suffices to evaluate $\vec{v}_W$ at the location of the particle. This
suggests that it is useful to define the quantity~\citep{felderhof2005JPhysChemB},
\begin{equation}
\label{eq:ReactionFieldTensor}
\tens{R}(\vec{r}_0;\omega) := \lim_{\vec{r} \to \vec{r}_0} \left [ \tens{G}(\vec{r}|\vec{r}_0;\omega) - \tens{G}^0(\vec{r} - \vec{r}_0;\omega) \right ],
\end{equation}
which Felderhof aptly calls the \emph{reaction field tensor}.

\subsection{The generalised Fax\'{e}n theorem of Mazur and Bedeaux}
\label{sec:faxen_theorem}

Felderhof's point-particle framework approaches the problem of determining the drag on a body in the presence of walls by using the formalism of
section~\ref{sec:felderhof_framework} to calculate the flow generated by a point force in the geometry, and later supposing that this flow be a background flow in which the
body is immersed. In order to determine the drag experienced by a spherical body suspended in a pre-existing flow, one needs to first calculate the change in the flow produced
by the presence of the body by applying the appropriate boundary conditions on the body, and then calculate the drag force experienced by the body. Generalised Fax\'{e}n
theorems provide a simple way to achieve this.

The formula for the drag on a stationary rigid sphere suspended in a pre-existing steady background flow $\vec{v}_0(\vec{r})$ was first derived by~\citet{faxen1921thesis}. The drag force is given by a very simple formula -- for no-slip boundary conditions on the sphere, $\vec{F}^\mathrm{drag} = \gamma_s \vec{\bar{v}}_0^S$, where $\vec{\bar{v}}_0^S$ is the
average of the background flow field over the surface of the sphere, and $\gamma_s = 6 \thepi \eta a$ is the well-known steady Stokes drag coefficient.

The~\citet{faxen1921thesis} theorem has been generalised to obtain the drag force on a sphere with a no-slip boundary in incompressible~\citep{mazur1974faxen} and
compressible~\citep{bedeaux1974generalization} unsteady Stokes flow.~\citet{albano1975faxen} have generalised the incompressible version to the case of partial slip boundary
conditions on the sphere, and generalisation to the force density induced on the sphere has been effected by~\citet{felderhof1976faxen}. We review here, the incompressible
case for translational oscillations of a no-slip sphere derived by~\citet{mazur1974faxen}.

Consider an arbitrary background fluid flow described by $\{\vec{v}_0(\vec{r}; \omega)$, $P_0(\vec{r}; \omega)\}$ extant in $\mathbb{R}^3$, which solves the unsteady
incompressible Stokes equations with a body force distribution $\vec{S}_0(\vec{r}; \omega)$ consistent with the background flow, i.e.
\begin{equation}
\label{eq:UnsteadyStokesFourierWithSource}
\begin{aligned}
\Delta \vec{v}_0 -\alpha^2 \vec{v}_0 &= \frac{\bnabla P_0 - \vec{S}_0}{\eta}, \\
\bnabla \bcdot \vec{v}_0 &= 0.
\end{aligned}
\end{equation}
Suppose that we now place a no-slip sphere of radius $a$, which executes small translational oscillations with velocity $\vec{u}_\omega$ in the fluid under the influence of
some external force. The fluid flow is altered by the boundary conditions imposed by the sphere. Since the system is linear, we could think of this as being due to an
additional flow $\{\vec{v}'(\vec{r}; \omega)$, $P'(\vec{r}; \omega)\}$. Once again due to linearity, we expect that this flow depends linearly on both the boundary condition
$\vec{u}_\omega$ and the background flow $\vec{v}_0$, $P_0$.

This relationship is expressed readily if we convert the boundary condition into a source, as is often done in electrodynamics and fluid mechanics. Introducing an induced
force\footnote{The notion of induced forces, as described by~\citet{mazur1974faxen}, is analogous to the notion of bound charges in electrostatics.} density $\vec{S}_\mathrm{ind}(\vec{r}; \omega)$ that has support only in the region occupied by the sphere (which we shall assume in this section to be
$|\vec{r}| \leq a$), we obtain the equations
\begin{equation}
\label{eq:UnsteadyStokesForSpherePerturbation}
\begin{aligned}
\Delta \vec{v}' -\alpha^2 \vec{v}' &= \frac{\bnabla P' - \vec{S}_{\mathrm{ind}}}{\eta}, \\
\bnabla \bcdot \vec{v}' &= 0.
\end{aligned}
\end{equation}
In the above, we assume that there is no longer a boundary, but fluid filling the region $|\vec{r}| \leq a$. A key requirement is that $\vec{S}_{\mathrm{ind}}$ be chosen so the momentum flux through the boundary in this problem matches that through the sphere oscillating with velocity $\vec{u}_\omega$. We shall additionally require that the total flow $\vec{v} = \vec{v}_0 + \vec{v}'$ be equal to $\vec{u}_\omega$ in the entire $|\vec{r}| \leq a$ region.

We may write the formal solution of~\eqref{eq:UnsteadyStokesForSpherePerturbation} as\footnote{Equation~\eqref{eq:FormalSolutionForFullFlow} can be seen to be identical to equation (3.15) of~\citet{mazur1974faxen} upon employing~\eqref{eq:UnsteadyOseenTensor}.}
\begin{equation}
\label{eq:FormalSolutionForFullFlow}
\vec{v}(\vec{r}; \omega) = \vec{v}_0(\vec{r}; \omega) + \int_{|\vec{r'}| \leq a} \mathrm{d}^3r' \; \tens{G}^0(\vec{r} - \vec{r}'; \omega) \bcdot \vec{S}_\mathrm{ind}(\vec{r}'; \omega)
\end{equation}
To find $\vec{S}_\mathrm{ind}$, it appears that one would need to solve the above integral equation, where the left hand side is known to be $\vec{u}_\omega$ inside the spherical region.
However, it turns out its explicit value is not required for our purposes -- to compute the drag force $\dragforce$ on the sphere, it suffices to compute the integrated
value of $\vec{S}_\mathrm{ind}$ over the volume of the sphere, for
\begin{equation}
\begin{aligned}
\label{eq:InducedForceAndDrag}
\dragforce &= \oint_{|\vec{r}| = a} \mathrm{d}^2r \; \pmb{\sigma} \bcdot \hat{n} =  \int_{|\vec{r}| \leq a} \mathrm{d}^3r \; \bnabla \bcdot \pmb{\sigma}\\
&= - \left [ \imagunit \omega m_f \vec{u}_\omega + \int_{|\vec{r}| \leq a} \mathrm{d}^3 r \; \vec{S}_\mathrm{ind}(\vec{r}; \omega) \right ]
\end{aligned}
\end{equation}
as required for the induced force to mimic the presence of the sphere, with $m_f = \frac{4}{3} \thepi a^3 \rho_f$ being the mass of fluid displaced by the sphere. The last step was effected by writing
\begin{equation}
\bnabla \bcdot \pmb{\sigma} = -\bnabla P + \eta \Delta \vec{v}  =  \eta \alpha^2 \vec{v} - \vec{S}_{\mathrm{ind}} - \vec{S}_0,
\end{equation}
and noting that $\vec{S}_0$ may be set without loss of generality to 0 in the region $r \leq a$ (by lumping whatever value it had into $\vec{S}_\mathrm{ind}$), as well as that $\vec{v}$ has the constant value $\vec{u}_\omega$ in the region $r \leq a$ whereby the integral of $\vec{v}$ over that region is simply $\vec{u}_\omega$ times the volume of the sphere.

By various manipulations, it can be shown that averages of the equation~\eqref{eq:FormalSolutionForFullFlow} suffice to determine $\int_{|\vec{r}| \leq a} \mathrm{d}^3 r \; \vec{S}_\mathrm{ind}$, whereby setting $\vec{v}(\vec{r}; \omega) = \vec{u}_\omega$ for $|\vec{r}| \leq a$ and averaging~\eqref{eq:FormalSolutionForFullFlow} over the surface and the volume of the sphere, the desired result for the drag force is obtained to be
\begin{equation}
\label{eq:Faxen}
\dragforce = -\gamma_0(\omega) \vec{u}_\omega + \gamma_s \left [ (1 + \alpha a) \vec{\bar{v}}_0^S(\omega) + \frac{1}{3} \alpha^2 a^2 \vec{\bar{v}}_0^V(\omega) \right ],
\end{equation}
where
\begin{equation}
\label{eq:StokesBoussinesq}
\gamma_0(\omega) := \gamma_s \left ( 1 + \alpha a + \frac{1}{9}\alpha^2 a^2 \right )
\end{equation}
is the unsteady Stokes-Boussinesq drag coefficient for a sphere, and $\vec{\bar{v}}_0^S$ and $\vec{\bar{v}}_0^V$ denote the averages of $\vec{v}_0$ over the surface and volume of the
sphere respectively. The above result is the generalisation of Fax\'{e}n's theorem by~\citet{mazur1974faxen}.

In the point-particle framework of Felderhof, the flow $\vec{v}_W$ calculated using~\eqref{eq:ReflectedFlowField} is considered to be the
background flow $\vec{v}_0$. In addition, the surface and volume averages of $\vec{v}_0$ are approximated by evaluating $\vec{v}_W$ at the centre of the
sphere. Thus, using the definition~\eqref{eq:ReactionFieldTensor} of the reaction field tensor, we obtain in the point-particle limit,
\begin{equation}
\label{eq:FaxenPointParticle}
\dragforce = -\gamma_0(\omega) \vec{u}_\omega + \gamma_s \left (1 + \delta + \frac{\delta^2}{3}  \right ) \tens{R} \bcdot \indforce,
\end{equation}
where we have introduced the notation $\delta := \alpha a$. We must note that in the adaptation of the generalised Fax\'{e}n theorem to Felderhof's framework, the net flow $\vec{v} = \vec{v}_W + \vec{v}'$ does not necessarily satisfy boundary conditions on the walls, and this is part of the approximation.

\subsection{The appropriate choice of the point force $\indforce$}
\label{sec:induced_force}

We now wish to address the following question: what must the point force $\indforce$ of section~\ref{sec:felderhof_framework} be, to capture the effects on the fluid due to
the presence of the body $S$?~\citet[][eq. (2.8)]{felderhof2005JPhysChemB} uses the external force $\extforce$ that acts on the body by means of some external agent to keep it oscillating with velocity
$\vec{u}_\omega$. However, as some of the momentum delivered by the force $\extforce$ goes into accelerating the body $S$, it is unlikely that this is equal to the force
applied on the fluid. It seems reasonable that the force must reproduce the momentum transport through the boundary $\partial S$ of the small body, when the body's volume is
replaced by fluid. This is the notion of induced force of~\citet{mazur1974faxen}, which as we described in section~\ref{sec:faxen_theorem}, can be used to replace boundary
conditions by sources.

In the previous section, we stated in equation~\eqref{eq:InducedForceAndDrag} an expression for the total induced force that replaces a spherical boundary oscillating at
$\vec{u}_\omega$. Based on that, we propose that the value of the point force must be given by the same net force concentrated at a point,
\begin{equation}
\label{eq:InducedForce}
\indforce = -\dragforce - \imagunit \omega m_f \vec{u}_\omega,
\end{equation}
possibly also for bodies of generic shape. In this equation, we note that no reference has been made to the properties of the  body or the external force acting on it. These aspects, however, do affect the velocity $\vec{u}_\omega$ through the equation of motion of the body,
\begin{equation}
\label{eq:EquationOfMotionBody}
\extforce = -\dragforce - \imagunit \omega m_p \vec{u}_\omega,
\end{equation}
which leads to the alternate expression for the net induced force $\indforce$,
\begin{equation}
\indforce = \extforce + \imagunit \omega (m_p - m_f) \vec{u}_\omega.
\end{equation}
as used by~\citet{felderhof2005JPhysChemB}. We would recover Felderhof's proposal of using $\extforce$ as the force that represents the body if the body had the same density as the fluid.

To establish our proposal for $\indforce$, we observe that we may write the velocity field $\vec{v}$ produced by the oscillating body at an arbitrary point $\vec{r}$ using the Green's function of equation~\eqref{eq:StokesletProblemMomentum} as
\begin{equation}
\vec{v}(\vec{r}; \omega) = \int_{S} \mathrm{d}^3 r' \; \tens{G}(\vec{r}|\vec{r}';\omega) \bcdot \vec{S}_{\mathrm{ind}}(\vec{r}'; \omega),
\end{equation}
where we have replaced the body $S$ by an appropriate induced force density. As is typical of multipole expansions, we may expand $\tens{G}$ in the source point in the far-field limit (i.e. $|\vec{r}| \gg 1/|\alpha|,\: L$ where $L$ denotes the size of the body) to obtain
\begin{equation}
\vec{v}(\vec{r}; \omega) = \int_{S} \mathrm{d}^3 r' \; \left [ \tens{G}(\vec{r}|\vec{r}_0;\omega) + (\vec{r} - \vec{r}_0) \bcdot \bnabla \tens{G}(\vec{r}|\vec{r}_0;\omega) + \ldots \right ] \bcdot \vec{S}_{\mathrm{ind}}(\vec{r}'; \omega),
\end{equation}
where $\vec{r}_0$ is some notion of the centre of the body. Truncating the expansion to the first term gives the expression for the velocity due to a point force at $\vec{r}_0$, whose strength is indeed given by
\begin{equation}
\indforce = \int_{S} \mathrm{d}^3 r' \; \vec{S}_{\mathrm{ind}}(\vec{r}'; \omega).
\end{equation}

We further ratify our result for $\indforce$ by checking it for the case of unbounded spherical bodies in the following manner: we take the far-field limit (i.e.
$|\vec{r}| \gg 1/|\alpha|,\: a$) of the solution for the flow $\vec{v}_{\omega}^{\mathrm{S}}(r,\theta)$ produced by a sphere of radius $a$ at the origin oscillating with
velocity $\vec{u}_\omega$ \citep[see e.g.][\S 24]{landaulifshitz1987fluid}, and compare it against the flow $\vec{v}^{\mathrm{PF}}_{\omega}$ generated by a point force
$\vec{F}_\omega$ at the origin \citep[see e.g.][\S 6.2]{kim2013microhydrodynamics}. For conciseness, we compare only the radial component.

Using spherical polar coordinates with the polar axis along $\vec{u}_\omega$, and introducing the notation $\varepsilon := a/r$, we
find that the radial component of the velocity field for a sphere is given by
\begin{equation}
\hat{e}_r \bcdot \vec{v}^{\mathrm{S}}_{\omega}(r,\theta) = -u_\omega \frac{2f'(r)}{r} \cos \theta,
\end{equation}
where~\citep[as given by][\S 24, Prob. 5]{landaulifshitz1987fluid}
\begin{equation}
\label{eq:FPrimeOverRFarFieldForm}
\frac{f'(r)}{r} = \frac{3 \varepsilon^3}{2 \delta^2} \left [ \eulere^{\delta(1-1/\varepsilon)}\left ( 1 + \frac{\delta}{\varepsilon} \right) - \left (
  1 + \delta + \frac{\delta^2}{3} \right ) \right ].
\end{equation}
On the other hand, for an unsteady stokeslet of strength $\vec{F}_\omega = F_\omega \hat{e}_z$, where $\hat{e}_z$ is the unit vector along the polar axis, we have
\begin{equation}
  \hat{e}_r \bcdot \vec{v}^{\mathrm{PF}}_{\omega} = \frac{2 \alpha \varepsilon^3}{\delta^3} \left [ 1 - \left ( 1 + \frac{\delta}{\varepsilon} \right )
    \eulere^{-\delta/\varepsilon} \right ] \frac{F_\omega}{4\thepi \eta} \cos \theta.
\end{equation}
In the far-field limit ($\varepsilon \to 0^+$ with $\delta$ fixed and finite), we may drop the subdominant exponential terms of the form $\eulere^{-\delta / \varepsilon}$ and
obtain
\begin{equation}
\begin{aligned}
  \frac{f'(r)}{r} &\sim -\frac{3\varepsilon^3}{2\delta^2} \left ( 1 + \delta + \frac{\delta^2}{3} \right ), \\
  \hat{e}_r\bcdot \vec{v}_{\omega}^{\mathrm{S}} &\sim u_\omega \cos \theta \; \frac{3 \varepsilon^3}{\delta^2} \left ( 1 + \delta + \frac{\delta^2}{3} \right ), \\
  \hat{e}_r\bcdot \vec{v}_{\omega}^{\mathrm{PF}} &\sim \frac{4\alpha\varepsilon^3}{\delta^3}\frac{F_\omega}{8\thepi\eta} \cos \theta. \\
\end{aligned}
\end{equation}
By setting the latter two expressions equal to each other, we find that,
\begin{equation}
\begin{aligned}
F_\omega &= \gamma_s u_\omega \left(1 + \delta + \frac{\delta^2}{3} \right) \\
&= \gamma_s u_\omega \left (1 + \delta + \frac{\delta^2}{9} \right) - \imagunit \omega m_f u_\omega.
\end{aligned}
\end{equation}
We now identify the first term to be $-F^\mathrm{drag}_\omega = \gamma_0(\omega) u_\omega$, whereby we find that $\vec{F}_\omega$ is indeed equal to the induced force $\indforce$. We are
hence led to conclude that an unsteady stokeslet of strength $\indforce$ as defined by equation~\eqref{eq:InducedForce} reproduces the far-field behaviour of a sphere, which would not be the case for Felderhof's choice of the external force $\extforce$. It is not unreasonable to expect from the physical and mathematical arguments presented earlier, that~\eqref{eq:InducedForce} also holds for bodies of generic shape.

\subsection{From the reaction field tensor to the dynamics of a sphere}
\label{sec:consequences_correction}

We will now follow Felderhof's approach, except with the modified point force $\indforce$ given by~\eqref{eq:InducedForce}, to arrive at expressions for the drag coefficient
and other relevant quantities characterising the dynamics of a \emph{sphere} oscillating in a fluid, in terms of the reaction field tensor $\tens{R}$.

We start by using equations~\eqref{eq:FaxenPointParticle} and~\eqref{eq:InducedForce} to obtain
\begin{equation}
\label{eq:SelfReferentialInducedForce}
\indforce = \gamma_s \left ( 1 + \delta + \frac{\delta^2}{3} \right ) \left [ \vec{u}_\omega - \tens{R} \bcdot \indforce \right ],
\end{equation}
where we have used $-\imagunit \omega m_f = (2/9) \gamma_s \delta^2$ to simplify the expression. The difference between this expression and that
of~\citet[eq. 2.11]{felderhof2005JPhysChemB} is the use of $\indforce$ instead of $\extforce$. We may use this to solve for $\indforce$ as,
\begin{equation}
\label{eq:InducedForceFromReactionFieldTensor}
\indforce = \tilde{\gamma}_0(\omega) \left [ \mathds{1} + \tilde{\gamma}_0(\omega) \tens{R} \right ]^{-1} \bcdot \vec{u}_\omega,
\end{equation}
where we have defined for convenience,
\begin{equation}
\label{eq:FreeSpaceGammaTildeDefinition}
\tilde{\gamma}_0(\omega) := \gamma_s \left ( 1 + \delta + \frac{\delta^2}{3} \right ) = \gamma_0(\omega) - \imagunit \omega m_f.
\end{equation}

Thereafter, using the definition~\eqref{eq:DragCoefficientTensorDefinition} and plugging~\eqref{eq:InducedForceFromReactionFieldTensor} into~\eqref{eq:InducedForce}, we obtain
the drag coefficient tensor,
\begin{equation}
\label{eq:DragCoefficientTensorFromReactionFieldTensor}
\dragtensor = \imagunit \omega m_f \mathds{1} + \tilde{\gamma}_0(\omega) \left [ \mathds{1} + \tilde{\gamma}_0(\omega) \tens{R} \right ]^{-1}.
\end{equation}
Observe that if we define $\tens{\boldsymbol{\tilde{\gamma}}(\omega)} := \dragtensor - \imagunit \omega m_f \mathds{1}$ as before (so that
$\indforce = \tens{\boldsymbol{\tilde{\gamma}}}(\omega) \bcdot \vec{u}_\omega$), the correction of $\tilde{\gamma}_0$ to $\tens{\boldsymbol{\tilde{\gamma}}}$ through
$\tens{R}$ has the natural form of a Pad\'{e} approximant.

The mechanical admittance tensor $\admittance$, characterising the linear response of the velocity $\vec{u}_\omega$ of the sphere to the external force $\extforce$ acting on it, is
defined through
\begin{equation}
\vec{u}_\omega = \admittance \bcdot \extforce.
\end{equation}
It can be related to the drag coefficient through the equation of motion of the sphere~\eqref{eq:EquationOfMotionBody}, to obtain
\begin{equation}
\begin{aligned}
\label{eq:AdmittanceWithoutTrap}
\admittance &= \left [ -\imagunit \omega m_p \mathds{1} + \dragtensor \right ]^{-1} \\
            &= \left [ \tilde{\gamma}_0 \left (\mathds{1} + \tilde{\gamma}_0 \tens{R} \right )^{-1} - \imagunit \omega (m_p - m_f) \mathds{1} \right ]^{-1}.
\end{aligned}
\end{equation}
It is practically useful to include the effects of a harmonic restoring force $- \tens{K} \bcdot ( \vec{u}_\omega / (-\imagunit\omega ) ) $ in the equation of
motion~\citep{franosch2009persistent} (see also \S \ref{sec:brownian_motion}). The admittance then takes the form,
\begin{equation}
\label{eq:AdmittanceWithTrap}
\admittance = \left [ -\imagunit \omega m_p \mathds{1} + \dragtensor + \frac{\tens{K}}{-\imagunit \omega} \right ]^{-1}.
\end{equation}

\section{The validity of the point-particle approximation}
\label{sec:validity}

\subsection{Is the point-particle approximation valid?}

As we have stated earlier, there are two length scales in the problem in addition to the particle size -- the scale of the dimensions of the confining geometry $h$, and the
scale of the skin-depth of vorticity $1/|\alpha|$. The point-particle approximation neglects the size of the particle $a$ in comparison to both these length scales insofar as
the computation of the effect of the wall is concerned, and when computing the surface and volume averages of the flow that enter the generalised Fax\'{e}n theorem. It must be
noted that no approximations\footnote{excepting for the previously stated assumption that the boundary conditions may be applied on the equilibrium boundary of the
  sphere~\citep{mazur1974faxen}} are made in the generalised Fax\'{e}n theorem~\eqref{eq:Faxen} itself when the body is a sphere. However, for sufficiently large frequency
$\omega$ of oscillations, $1/|\alpha|$ can become comparable to $a$. This brings up the question of whether the point-particle approximation works at high
frequencies.

However, the agreement with experiment~\citep{mo2015pre} at frequencies $\omega \sim \eta / (\rho_f a^2)$ is very good. We explain this intuitively as follows: at these frequencies, the
vorticity shed by the boundaries has a very small skin-depth $1/|\alpha| \ll h$ and hence the vorticity from the wall is suppressed exponentially, and the reflected flow field
is well approximated by potential flow. Since the potential satisfies Laplace's equation, the multipole expansion and therefore the point-particle approximation works well. At
low frequencies $\omega \ll \eta / (\rho_f a^2)$, $1/|\alpha|$ is indeed large compared to $a$ and the approximation works as expected.

In order to harden the above argument, we shall set up a general formalism (section~\ref{sec:formal_framework}) for analysing the problem in terms of boundary integral equations, and then systematically delineate the approximations made in order to recover Felderhof's framework in section~\ref{sec:formalization}. The question then boils down to the validity of a far-field expansion of the unsteady Oseen tensor over a wide-range of frequencies, which we provide an argument for in section~\ref{sec:greens_function}. In section~\ref{sec:all_orders} we shall extend the perturbative calculation to higher orders and recover the Pad\'{e}-like form for the drag coefficient~\eqref{eq:DragCoefficientTensorFromReactionFieldTensor}.

\subsection{General formalism of boundary integral equations}
\label{sec:formal_framework}
In this sub-section, we cast our problem in the general formalism of boundary integral equations~\citep[see e.g.][]{pozrikidis1992}. In this and the following sub-sections, we drop explicit reference to $\omega$, the frequency, for notational simplicity. As before, the linearity and time-translation invariance ensure that the individual frequency components may be treated separately. The walls will be assumed to be larger in size than the distance from the particle to any of them. We also assume no-slip boundary conditions on all interfaces for the purposes of this discussion.

\newcommand{\rv}{\vec{r}}
\newcommand{\dthreer}{\mathrm{d}^3r}
\newcommand{\dtwor}{\mathrm{d}^2r}
\newcommand{\GO}[2]{\tens{G}^0({#1} - {#2})}
\newcommand{\GW}[2]{\tens{G}({#1} | {#2})}
\newcommand{\wallterm}[2]{\int_W \dtwor'_W \; \GO{{#2}}{\rv'_W} \bcdot \vec{S}_W^{({#1})}(\rv'_W)}
\newcommand{\bodyterm}[2]{\int_S \dthreer'_S \; \GO{{#2}}{\rv'_S} \bcdot \vec{S}_S^{({#1})}(\rv'_S)}
\newcommand{\formalexpansion}[2]{{{#1}_{#2}} = {{#1}^{(0)}_{#2}} + \lambda {{#1}^{(1)}_{#2}} + \lambda^2 {{#1}^{(2)}_{#2}} + \ldots}
\newcommand{\eqnsecondlineindent}{\qquad \qquad}

The problem at hand may be restated as follows: Find the drag force
\begin{equation}
\dragforcesubscript = - \left [ \imagunit \omega m_f \vec{u} + \int_{S} \dthreer'_S \; \vec{S}_S(\rv'_S) \right ]
\end{equation}
exerted on the surface of the particle $S$ oscillating with velocity $\vec{u}$, by the velocity field
\begin{equation}
\label{eq:FullVelocityEquation}
\vec{v}(\rv) = \int_S \dthreer'_S \; \GO{\rv}{\rv'_S} \bcdot \vec{S}_S(\rv'_S)
+ \int_W \dtwor'_W \; \GO{\rv}{\rv'_W} \bcdot \vec{S}_W(\rv'_W)
\end{equation}
which is assumed to be generated from two induced force distributions\footnote{We assume that the surfaces involved satisfy the requirements outlined by~\citet[\S 4.1, 4.2]{pozrikidis1992} for representation of the flow by a single-layer potential, i.e. the surfaces are Lyapunov surfaces. While the integral condition $\int_D \vec{v}(\rv') \bcdot \hat{n}(\rv') \; \dtwor' = 0$ is satisfied for compact $D$ by virtue of non-penetration, it can be shown to hold for each non-zero frequency component of the unsteady Stokes flow for an infinite wall too -- the flow generated from any finite force distribution decays sufficiently fast so that the flux through an infinitely large hemispherical surface is zero. In particular, one may explicitly solve for the Green's function satisfying no-slip conditions on a plane wall by means of a single-layer potential in the place of the wall.} -- a volume force density $\vec{S}_S$ supported in the volume (inclusive of the surface) of the body $S$,
and a surface force density\footnote{While it would be possible to use a volume force density instead here as well, it does not make a difference for our purposes.}
$\vec{S}_W$ supported on the surfaces of the walls $W = \bigcup_{i} W_i$ -- which are to be determined from the no-slip boundary conditions. Thus $\vec{S}_S$ and $\vec{S}_W$ satisfy the Fredholm integral equations of the first kind,
\begin{equation}
\label{eq:FullBodyIntegralEquation}
\vec{u} = \int_S \dthreer'_S \; \GO{\rv_S}{\rv'_S} \bcdot \vec{S}_S(\rv'_S)
+ \int_W \dtwor'_W \; \GO{\rv_S}{\rv'_W} \bcdot \vec{S}_W(\rv'_W) \qquad \forall \rv_S \in S,
\end{equation}
\begin{equation}
\begin{aligned}[b]
\label{eq:FullWallIntegralEquation}
0 = \int_S \dthreer'_S \; \GO{\rv_W}{\rv'_S} &\bcdot \vec{S}_S(\rv'_S)\\
&+ \int_W \dtwor'_W \; \GO{\rv_W}{\rv'_W} \bcdot \vec{S}_W(\rv'_W) \qquad \forall \rv_W \in W.
\end{aligned}
\end{equation}
We remark that if the Green's function $\tens{G}$ that satisfies the boundary conditions on the walls were known, it would be possible to rewrite the problem purely in terms of $S_S$ as
\begin{equation}
\begin{aligned}
\vec{v}(\rv) &= \int_S \dthreer'_S \; \GW{\rv}{\rv'_S} \bcdot \vec{S}_S(\rv'_S),\\
\vec{u} &= \int_S \dthreer'_S \; \GW{\rv_S}{\rv'_S} \bcdot \vec{S}_S(\rv'_S) \qquad \forall \rv_S \in S.
\end{aligned}
\end{equation}

We now proceed to introduce a formal perturbative expansion in a parameter $\lambda$, which represents the ratio of the body size ($\sim a$) to the distance to the walls
($\sim h$). We begin by introducing expansions for the force distributions,
\begin{equation}
\begin{gathered}
  \label{eq:SourceExpansions}
  \formalexpansion{\vec{S}}{S},\\
  \formalexpansion{\vec{S}}{W}.
\end{gathered}
\end{equation}
These expansions induce expansions for the other quantities in the problem,
\begin{equation}
\begin{gathered}
  \label{eq:OtherExpansions}
  \formalexpansion{\vec{v}}{},\\
  \formalexpansion{\vec{F}}{\mathrm{drag}}.
\end{gathered}
\end{equation}

In analogy with examples from electrostatics, we expect that the effect of the induced force $\vec{S}_W$ on the walls is diminished in the region occupied by the body. We
shall further investigate this assumption, restated formally in~\eqref{eq:ExpansionAssumption}, at the end of this section. To emphasise this, we
rewrite~\eqref{eq:FullBodyIntegralEquation} as
\begin{equation}
\begin{aligned}[b]
\label{eq:FullBodyIntegralEquationWithFormalParameter}
\vec{u} =& \int_S \dthreer'_S \; \GO{\rv_S}{\rv'_S} \bcdot \vec{S}_S(\rv'_S)\\
&\eqnsecondlineindent + \lambda \int_W \dtwor'_W \; \frac{\GO{\rv_S}{\rv'_W} \bcdot \vec{S}_W(\rv'_W)}{\lambda} \qquad \forall \rv_S \in S.
\end{aligned}
\end{equation}
We would like a scheme where the velocity field from any $O(\lambda^k)$ truncation of the problem is faithful both near the walls and the body. The above convention makes this manifest.

We may now plug in the expansions and rewrite the problem~\{\eqref{eq:FullVelocityEquation},~\eqref{eq:FullWallIntegralEquation},~\eqref{eq:FullBodyIntegralEquationWithFormalParameter}\} order-by-order as
\begin{equation}
\label{eq:VelocityFromSource}
\vec{v}^{(n)}(\rv) = \bodyterm{n}{\rv} + \wallterm{n}{\rv},
\end{equation}
with the boundary condition on the body $S$
\begin{equation}
\begin{aligned}[c]
\label{eq:BodyIntegralEquation}
\vec{u} = &\bodyterm{0}{\rv_S} \qquad \forall \rv_S \in S,\\
0 = &\bodyterm{n+1}{\rv_S}\\
&\eqnsecondlineindent + \frac{1}{\lambda} \wallterm{n}{\rv_S} \qquad \forall n \geq 0,\;\forall \rv_S \in S,
\end{aligned}
\end{equation}
and the boundary condition on the walls
\begin{equation}
\begin{aligned}
\label{eq:WallIntegralEquation}
0 = &\bodyterm{n}{\rv_W}\\
&\eqnsecondlineindent + \wallterm{n}{\rv_W} \qquad \forall n \geq 0,\;\forall \rv_W \in W.
\end{aligned}
\end{equation}

We now proceed to investigate the assumption that
\begin{equation}
  \label{eq:ExpansionAssumption}
  \frac{1}{u} \int_W \dtwor'_W \; \GO{\rv_S}{\rv'_W} \bcdot \left [ \lambda^{k} \vec{S}_W^{(k)}(\rv'_W) \right ] \in O(\lambda^{k+1}).
\end{equation}
First, we note that $\vec{S}_W^{(k)}$ is obtained by solving~\eqref{eq:WallIntegralEquation} with the knowledge of $\vec{S}_S^{(k)}$. In the spirit of multipole expansions, since the free-space Green's function $\GO{\rv_W}{\rv'_S}$ in
the first integral of~\eqref{eq:WallIntegralEquation} is evaluated at a far separation, we may expand it in the vicinity of the location of the body $\rv_0$,
\begin{equation}
\begin{aligned}
\label{eq:QuestionableGreensFunctionExpansion}
\GO{\rv_W}{\rv'_S} &= \GO{\rv_W}{\rv_0} + (\rv'_S - \rv_0) \bcdot \bnabla \GO{\rv_W}{\rv_0} + \ldots\\
&= \GO{\rv_W}{\rv_0} + o(\lambda), \qquad \rv'_S \in S.
\end{aligned}
\end{equation}
The issue of the validity of such an expansion is subtle and will be addressed in detail in section~\ref{sec:greens_function}. Using this expansion
in~\eqref{eq:WallIntegralEquation}, we have
\begin{equation}
  \label{eq:WallIntegralEquationWithFarFieldExpansion}
  \wallterm{k}{\rv_W} = -\GO{\rv_W}{\rv_0} \bcdot \int_S \dthreer'_S \; \vec{S}_S^{(k)}(\rv'_S) + o(\lambda).
\end{equation}

We now state a useful result: if $\vec{\tilde{S}}_W$ satisfies the integral equation
\begin{equation}
  \label{eq:WallGreensFunctionProblem}
  \int_W \dtwor'_W \; \GO{\rv_W}{\rv'_W} \bcdot \vec{\tilde{S}}_W(\rv'_W) = - \GO{\rv_W}{\rv_0} \bcdot \vec{\tilde{F}} \quad \forall \rv_W \in W
\end{equation}
for arbitrary point $\rv_0$ and force $\vec{\tilde{F}}$, then for general $\rv$ in the domain,
\begin{equation}
  \label{eq:WallGreensFunctionSolution}
  \int_W \dtwor'_W \GO{\rv}{\rv'_W} \bcdot \vec{\tilde{S}}_W(\rv'_W) = \left [ \tens{G}(\rv|\rv_0) - \GO{\rv}{\rv_0} \right ] \bcdot \vec{\tilde{F}},
\end{equation}
where $\tens{G}(\rv|\rv_0)$ is the Green's function that satisfies no-slip boundary conditions on the walls. This is easily seen if we set up the problem for the no-slip Green's
function for the walls by imposing the boundary condition through a surface force distribution $\vec{\tilde{S}}_W$ on the walls.

If we choose for $\vec{\tilde{F}}$ the force
\begin{equation}
\int_S \dthreer'_S \; \vec{S}^{(k)}_S(\rv'_S),
\end{equation}
we find by comparing~\eqref{eq:WallIntegralEquationWithFarFieldExpansion} and~\eqref{eq:WallGreensFunctionProblem} that
\begin{equation}
  \begin{aligned}
    \label{eq:WallIntegralEquationSolution}
    &\wallterm{k}{\rv}\\
    &\eqnsecondlineindent = \left [ \tens{G}(\rv|\rv_0) - \GO{\rv}{\rv_0} \right ] \bcdot \int_S \dthreer'_S \; \vec{S}^{(k)}_S(\rv'_S) + o(\lambda)
  \end{aligned}
\end{equation}
for any point $\rv$ in the domain.

Finally, we observe that we may approximate the expression in question as
\begin{equation}
  \begin{aligned}
    \label{eq:ExpansionCheck}
    &\frac{\lambda^k}{u} \wallterm{k}{\rv_S}\\
    &\eqnsecondlineindent = \frac{\lambda^k}{u} \wallterm{k}{\rv_0} + o(\lambda^{k+1})\\
  &\eqnsecondlineindent = \frac{\lambda^k}{u} \lim_{\rv_S \to \rv_0} \left [ \tens{G}(\rv_S|\rv_0) - \GO{\rv_S}{\rv_0} \right ] \bcdot \int_S \dthreer'_S \; \vec{S}^{(k)}_S(\rv'_S) + o(\lambda^{k+1})\\
  &\eqnsecondlineindent = \frac{\lambda^k}{u} \tens{R}(\rv_0) \bcdot \int_S \dthreer'_S \; \vec{S}^{(k)}_S(\rv'_S) + o(\lambda^{k+1}).
\end{aligned}
\end{equation}
Thus, if
\begin{equation}
  \label{eq:ConditionOnReactionFieldTensor}
  \frac{1}{u} \tens{R}(\rv_0) \bcdot \int_S \dthreer'_S \; \vec{S}^{(k)}_S(\rv'_S) \in O(\lambda),
\end{equation}
then the assumption~\eqref{eq:ExpansionAssumption} holds. Intuitively, one may expect the above condition to hold on the grounds that the reaction field tensor is the reflected flow evaluated at the location of the particle, and this reflected flow must be suppressed at least as $1/h$, $h$ being the distance to the wall, whereas one would expect the remaining terms to produce a factor of $a$.

\subsection{Formalisation of the point-particle approximation}
\label{sec:formalization}

In this sub-section and the next, we seek to formalise the point-particle framework by explicitly performing all the approximations involved in a systematic manner, using the formalism developed in the previous sub-section. We shall eventually specialise $S$ to be a sphere while still keeping $W$ arbitrary.

To solve the problem at order $n = 0$, we begin by noting that the solution to the first of~\eqref{eq:BodyIntegralEquation}
is the induced force on the body oscillating with velocity $\vec{u}$ in unbounded fluid, whereby
\begin{equation}
  \label{eq:VelocityUnboundedBody}
\vec{v}^S(\rv) \equiv \bodyterm{0}{\rv}
\end{equation}
where we have used $\vec{v}^S(\rv)$ to denote the velocity field generated by the body $S$ oscillating in unbounded fluid. We must now find $\vec{S}^{(0)}_W$ using~\eqref{eq:WallIntegralEquation}, which is not analytically tractable without approximation. Therefore, we make the same approximations that lead to~\eqref{eq:WallIntegralEquationWithFarFieldExpansion}. As we will see shortly, to compute the drag force to first order, we do not need to know $\vec{S}^{(0)}_W$, but only need to be able to compute the effect of this distribution in the vicinity of the body. Proceeding as we did in section~\ref{sec:formal_framework}, we may therefore write~\eqref{eq:WallIntegralEquationSolution} for $k = 0$ as
\begin{equation}
\label{eq:IntegralRepForReflectedFlow}
\wallterm{0}{\rv} = \left [ \tens{G}(\rv|\rv_0) - \GO{\rv}{\rv_0} \right ] \bcdot \vec{F}^{(0)} + o(\lambda),
\end{equation}
where we have defined
\begin{equation}
\vec{F}^{(k)} := \int_S \dthreer'_S \; \vec{S}^{(k)}_S(\rv'_S).
\end{equation}

We may now write~\eqref{eq:BodyIntegralEquation} for $n = 1$ as
\begin{equation}
  \begin{aligned}
    \label{eq:FaxenProblemDisguised}
    0 = &\bodyterm{1}{\rv_S}\\
    &\qquad \qquad+ \frac{1}{\lambda} \left [ \tens{G}(\rv_S|\rv_0) - \GO{\rv_S}{\rv_0} \right ] \bcdot \vec{F}^{(0)} \quad \forall \rv_S \in S.
    \end{aligned}
\end{equation}

Our aim is to determine the correction to the drag $\dragforcesubscript^{(1)}$ resulting from the field $\vec{v}^{(1)}$. To determine the drag force, we only need the velocity in the near field of the body, whereby in~\eqref{eq:VelocityFromSource} at any order $n$, we may discard the contribution from $\vec{S}^{(n)}_W$, as the unsteady Oseen tensor multiplying it contributes an extra $O(\lambda)$ when compared to the contribution from the first term when the point of evaluation $\rv$ is close to the body. As a result, we obtain
\begin{equation}
  \label{eq:NearFieldVelocityAtOrderN}
\vec{v}^{(n)}(\rv_S) = \bodyterm{n}{\rv_S} + O(\lambda) \qquad \forall \rv_S \in S.
\end{equation}
Thus, we observe that at order 0, we may use $\vec{v}^S$ of~\eqref{eq:VelocityUnboundedBody} to compute $\dragforcesubscript^{(0)}$, which is simply equal to the drag force on the body oscillating in unbounded fluid; and at order 1, knowledge of $\vec{S}^{(1)}_W$ is not required for the computation of $\dragforcesubscript^{(1)}$.

We now specialise to $S$ being a sphere of radius $a$ and proceed to determine $\dragforcesubscript^{(1)}$ for this case. If we set
$\vec{v}_0(\rv) := \left [ \tens{G}(\rv|\rv_0) - \GO{\rv}{\rv_0} \right ] \bcdot \vec{F}^{(0)}$ and $\vec{v}(|\rv| \leq a) = 0$ in equation~\eqref{eq:FormalSolutionForFullFlow}, we see that~\eqref{eq:FaxenProblemDisguised} is identical to~\eqref{eq:FormalSolutionForFullFlow}. Therefore, $\dragforcesubscript^{(1)}$ is given by the generalised Fax\'{e}n theorem of equation~\eqref{eq:Faxen}, whereby we may write
\begin{equation}
\label{eq:FaxenRecoveredFormally}
\lambda \dragforcesubscript^{(1)} = \gamma_s \left [ (1 + \alpha a) \left < \vec{v}^{(1)}_W \right >_S + \frac{1}{3} \alpha^2 a^2 \left < \vec{v}^{(1)}_W \right >_V \right ]
\end{equation}
with
\begin{equation}
\label{eq:FirstOrderReactionField}
\lambda \vec{v}^{(1)}_W := \left [ \tens{G}(\rv|\rv_0) - \GO{\rv}{\rv_0} \right ] \bcdot \vec{F}^{(0)},
\end{equation}
and $\left < \; \right >_S$ and $\left < \; \right >_V$ denoting surface and volume averages over the sphere respectively.
We wish to note that the analysis shows that the reaction field tensor is already $O(\lambda)$, which may be verified with Felderhof's expressions for the case of a flat
wall. So the total drag force may be written by adding $\dragforcesubscript^{(0)} = -\gamma_0(\omega) \vec{u}$ and $\lambda \dragforcesubscript^{(1)}$ recovering~\eqref{eq:Faxen} for the drag up to
first order, with $\vec{v}_0 = \lambda \vec{v}^{(1)}_W$.

We now make the approximation of truncating the infinite series to first order, excluding $o(\lambda)$ terms. As a side effect, we observe that
\begin{equation}
\lambda \vec{v}^{(1)}_W = \left [ \tens{G}(\rv|\rv_0) - \GO{\rv}{\rv_0} \right ] \bcdot \left ( \vec{F}^{(0)} + \lambda \vec{F}^{(1)} \right ) + o(\lambda).
\end{equation}
Identifying the parentheses in the above equation with the total induced force to first order,
\begin{equation}
\indforce = \vec{F}^{(0)} + \lambda \vec{F}^{(1)},
\end{equation}
we have shown that $\lambda \vec{v}^{(1)}_W$ is identical with $\vec{v}_W$ of equation~\eqref{eq:ReflectedFlowField} to lowest order.

We now investigate the possibility of replacing the surface and volume averages of $\lambda \vec{v}^{(1)}_W$ by evaluation of~\eqref{eq:FirstOrderReactionField} at $\rv \to \rv_0$. Applying the expansion of~\eqref{eq:QuestionableGreensFunctionExpansion} to~\eqref{eq:IntegralRepForReflectedFlow} evaluated for $\rv \in S$ (as done in~\eqref{eq:ExpansionCheck}), we see that $\lambda \vec{v}^{(1)}_W$ may indeed be assumed to have $o(\lambda)$ variation over the region occupied by the sphere. This justifies replacing the averages in~\eqref{eq:FaxenRecoveredFormally} with $\lambda \vec{v}^{(1)}_W$ evaluated as $\rv \to \rv_0$, subject to the validity of the expansion of~\eqref{eq:QuestionableGreensFunctionExpansion}.

Finally, we remark that it can be verified by plotting the explicit formulae given by~\citet{felderhof2005JPhysChemB} (also see erratum~\citet{felderhof2005erratum}) over a wide range of frequencies, that the components of the reaction field tensor for the no-slip sphere-plane-wall configuration, non-dimensionalized by multiplication with $\gamma_s$, do not significantly exceed $1 (a/h)$. Thus, the reaction field tensor for this particular case satisfies~\eqref{eq:ConditionOnReactionFieldTensor} and therefore validates the assumption of~\eqref{eq:ExpansionAssumption} by the arguments made in section~\ref{sec:formal_framework}.

\subsection{Far-field expansion of the unsteady Oseen tensor}
\label{sec:greens_function}
In this section, we will address the validity of an expansion of the unsteady\footnote{In this work, we will frequently drop the adjective unsteady to simplify our language. Since our work primarily concerns unsteady flow, this should not cause confusion. We will explicitly specify so when we refer to the steady Oseen tensor.} Oseen tensor, of the kind described in~\eqref{eq:QuestionableGreensFunctionExpansion}.

It is natural to our original problem to non-dimensionalize the Oseen tensor by $\gamma_s = 6\thepi \eta a$, given that our notion of forces is best normalised by $\gamma_s u$ -- this results in $\dragforcesubscript^{(0)}$ being $O(1)$ in our book-keeping. However, the Oseen tensor $\tens{G}^0(\vec{q})$ is naturally a function of $\alpha \vec{q}$, whereby for this analysis, it will be convenient to normalise it by $1 / \alpha$ and write
\begin{equation}
\begin{aligned}
4 \thepi \tens{\bar{G}}^0(\alpha \vec{q}) := 4 \thepi \eta \tens{G}^{0}(\vec{q}) / \alpha &= \hat{e}_q \hat{e}_q \frac{2}{(\alpha q)^3} \left [ 1 - (1 + \alpha q) \eulere^{-\alpha q} \right ]\\
&+ \left ( \mathds{1} - \hat{e}_q\hat{e}_q \right ) \frac{1}{(\alpha q)^3} \left [ (1 + \alpha q + \alpha^2 q^2) \eulere^{-\alpha q} - 1 \right ],
\end{aligned}
\end{equation}
where $\hat{e}_q$ denotes the unit vector along $\vec{q}$. In expansions of the form of~\eqref{eq:QuestionableGreensFunctionExpansion}, we write $\vec{q} = \vec{q}_L + \vec{q}_S$ where $\vec{q}_L$ denotes a large displacement of $O(h)$ and $\vec{q}_S$ denotes a small displacement of $O(a)$. Typically, $\vec{q}_L$ is $\rv_0 - \rv_W'$ where $\rv_W'$ is some point on the wall and $\vec{q}_S$ is $\rv_S - \rv_0$ where $\rv_S$ is some point in the body S. We write
\begin{equation}
\tens{\bar{G}}^0(\alpha \vec{q}) = \tens{\bar{G}}^0(\alpha \vec{q}_L) + \alpha \vec{q}_S \bcdot \bnabla_{\alpha \vec{q}_S} \tens{\bar{G}}^0(\alpha \vec{q}_L) + o(\alpha \vec{q}_S),
\end{equation}
where $\bnabla_{\alpha \vec{q}_S}$ denotes a gradient with respect to the quantity $\alpha \vec{q}_S$.
Such an expansion may be expected to be valid whenever the function is sufficiently slowly varying for small changes in $\vec{q}_S$ (i.e. changes over the scale of the size of the body). However, for sufficiently high wavenumbers $\alpha$, it appears that oscillating terms of the nature $\eulere^{\imagunit \,\Imag(\alpha) \,\vec{q}_S}$ would vary very rapidly -- whereby care must be taken to analyse such an expansion. Specifically, for the Helmholtz Green's function $-\eulere^{\imagunit k q} / (4 \thepi \imagunit k q)$, such an expansion is strictly valid only if $|\vec{q}_S| \ll 1/k \text{ and } |\vec{q}_S| \ll |\vec{q}_L|$, as is often noted when considering multipole expansions for electromagnetic radiation~\citep[see e.g.][\S 9.1]{jackson1999electrodynamics}. However, in the Oseen tensor, the complex wavenumber $\alpha = i k$ has a positive real part which causes significant suppression of the exponentials at large values of $\alpha$, in comparison to the terms originating from the fundamental solution of the Laplace equation $G(\vec{q}; 0)$. Essentially, for large $\alpha$, the contribution from $G(\vec{q}; \omega)$ becomes subdominant, which results in the expansion once again being valid for large $\alpha$. However, it must be noted that the expansion may not work if the subdominant behaviour is of primary interest, as could be the case.

We may verify the above intuitive remarks about the expansion by considering the ratio of the first order term in the Taylor expansion to the zeroth order term. To get an order of magnitude estimate, we will treat the longitudinal and transverse components of $\tens{\bar{G}}^0$ separately, and specifically set $q_L = h$ and $q_S = a$. Then, the desired ratios have the form
\begin{equation}
\frac{a}{h} \frac{\eulere^{-\nu}( 3 + 3\nu + \nu^2)-3}{\eulere^{-\nu}(1 + \nu) - 1},
\end{equation}
and
\begin{equation}
\frac{a}{h} \frac{\eulere^{-\nu}( 3 + 3\nu + 2\nu^2 + \nu^3 ) - 3}{\eulere^{-\nu}(1 + \nu + \nu^2) - 1},
\end{equation}
respectively, where we have used the shorthand $\nu := \alpha h$. While $a / h$ is assumed to be small from the geometry of the problem, no assumptions can be made about $\alpha$. So we must check that the parts of the ratios that contain only $\nu$ remain $\lesssim$ 1. Noting that $\nu$ has the form $\frac{1 - \imagunit}{\sqrt{2}} |\nu|$ and plotting these parts against a large range of values of $|\nu|$ (or alternately, by analysis), we find that the real and imaginary parts of the above ratios are bounded and do not significantly exceed 1 throughout the range. This indicates that the approximation can be expected to work well for all values of $\alpha$ so long as $a/h$ is small.

Intuitively speaking, this seems to suggest that at high frequencies, the primary contribution to the correction of the drag on the particle due to the presence of walls comes from the pressure, rather than from vorticity diffusion. The skin-depth of the vorticity is then too small for the effects of vorticity diffusion from the wall to be significant at the location of the particle and vice versa. The effects of vorticity local to the particle and the wall themselves are however, important, and they are accounted for correctly in the framework.

Thus, we have shown that Felderhof's point-particle framework, with our modified point force $\indforce$, may be expected to work well at all frequencies\footnote{It must still be the case however, as stated earlier, that the frequencies be small enough that the fluid may be considered to be incompressible. For micron-sized particles in water, the regime where compressibility matters is usually of the order of GHz.} so long as
$a/h \ll 1$.

\subsection{Computing the perturbative expansion to all orders}
\label{sec:all_orders}
We begin by rewriting the result of the generalised Fax\'{e}n theorem (section~\ref{sec:faxen_theorem}) in a form that is readily usable in this section. In
equation~\eqref{eq:FormalSolutionForFullFlow}, we set $\vec{v} = 0$ within the region of the sphere, and we correspondingly set $\vec{u}_\omega = 0$ in~\eqref{eq:Faxen} and
use~\eqref{eq:InducedForceAndDrag} to obtain the following result: If the force distribution $\vec{\tilde{S}}$ on a sphere of radius $a$ obeys the integral equation
\begin{equation}
  \label{eq:FaxenReformulationProblem}
  \int_{\left | \rv_S' \right | \leq a} \GO{\rv_S}{\rv_S'} \bcdot \vec{\tilde{S}}(\rv_S') \; \mathrm{d}^3r'_S = -\vec{v}_0(\rv_S), \qquad \forall | \rv_S | \leq a,
\end{equation}
for some vector field $\vec{v}_0(\rv_S)$ having support in the region of the sphere, then we may write the net induced force in the region of the sphere as
\begin{equation}
  \label{eq:FaxenReformulationSolution}
  \int_{\left | \rv_S' \right | \leq a} \vec{\tilde{S}}(\rv_S') \; \mathrm{d}^3r'_S = - \gamma_s \left [ (1 + \alpha a) \vec{\bar{v}}_0^S(\omega) + \frac{1}{3} \alpha^2 a^2 \vec{\bar{v}}_0^V(\omega) \right ].
\end{equation}

We now consider the extension of the calculation performed in section~\ref{sec:formalization} to higher orders for the case where $S$ is a sphere of radius $a$. By using the result~\eqref{eq:ExpansionCheck} in the boundary condition on the body~\eqref{eq:BodyIntegralEquation}, we may write
\begin{equation}
  \begin{aligned}
    0 &= \bodyterm{n+1}{\rv_S}\\
    &\eqnsecondlineindent + \frac{1}{\lambda} \tens{R}(\rv_0) \bcdot \int_S \dthreer'_S \; \vec{S}^{(n)}_S(\rv'_S) \qquad \forall n \geq 0,\;\forall \rv_S \in S.
  \end{aligned}
\end{equation}
We note that the second term is independent of $\rv_S$ to the lowest order.

By comparing the above equation with equation~\eqref{eq:FaxenReformulationProblem}, we see that~\eqref{eq:FaxenReformulationSolution} gives us
\begin{equation}
  \int_S \dthreer'_S \; \vec{S}^{(n+1)}_S(\rv'_S) = \left [ \frac{-\tilde{\gamma}_0 \tens{R}(\rv_0)}{\lambda} \right ]^{n + 1} \bcdot \int_S \dthreer'_S \; \vec{S}^{(0)}_S(\rv'_S),
\end{equation}
which yields a geometric series. This indicates that we may write the net induced force on the sphere as
\begin{equation}
  \begin{aligned}
    \indforce = \int_S \dthreer'_S \; \vec{S}_S(\rv'_S) &= \sum_{k = 0}^\infty \int_S \dthreer'_S \; \lambda^k \vec{S}^{(k)}_S(\rv'_S)\\
    &= \left ( \sum_{k = 0}^\infty \left [ -\tilde{\gamma}_0 \tens{R}(\rv_0) \right ]^k \right ) \bcdot \int_S \dthreer'_S \; \vec{S}^{(0)}_S(\rv'_S)\\
    &= \left [ \mathds{1} + \tilde{\gamma}_0 \tens{R}(\rv_0) \right ]^{-1} \bcdot \int_S \dthreer'_S \; \vec{S}^{(0)}_S(\rv'_S),
  \end{aligned}
\end{equation}
provided the geometric series converges.

By comparing the first of~\eqref{eq:BodyIntegralEquation} with~\eqref{eq:FaxenReformulationProblem}, we find from~\eqref{eq:FaxenReformulationSolution} that
\begin{equation*}
  \int_S \dthreer'_S \; \vec{S}^{(0)}_S(\rv'_S) = \tilde{\gamma}_0 \vec{u}.
\end{equation*}
Thereafter, using~\eqref{eq:InducedForceAndDrag} we find that the drag force to all orders in $a/h$ is given by
\begin{equation}
  \dragforce = - \imagunit \omega m_f - \tilde{\gamma}_0 \left [ \mathds{1} + \tilde{\gamma}_0 \tens{R}(\rv_0) \right ]^{-1} \bcdot \vec{u},
\end{equation}
whereby we recover the result~\eqref{eq:DragCoefficientTensorFromReactionFieldTensor}. Thus, it appears that in the region of convergence of the geometric series, the results
of the point-particle framework are correct to all orders of perturbation theory.

However, this does not mean that it is exact irrespective of how large $a/h$ is, since the perturbative process does not necessarily capture the corrections that lie beyond
all orders faithfully, which become significant as $a/h \to 1$. In fact, in the next section, we will compare the first order results from the point-particle approximation
against the method of reflections for the simpler case of full-slip boundary conditions on the wall, and discover that the subdominant terms do differ.

\section{Method of Reflections -- a no-slip sphere near a full-slip plane wall}
\label{sec:reflection_method}

The method of reflections has been heavily used as an approximation method in the context of steady Stokes flow~\citep[see e.g.][]{happel1965,kim2013microhydrodynamics}. A
proof of the convergence of the iterative process for steady Stokes flows under certain restrictions exists~\citep{luke1989convergence}, although this has not been extended to
unsteady Stokes flows (to the best of our knowledge)\footnote{A formalism of the sort developed in~\ref{sec:formal_framework} could serve as a starting point for a proof.}. The method of reflections has been used in the context of unsteady Stokes flows for the case of two spheres with no-slip
boundary conditions by~\citet{ardekani2006Reflections}, but their procedure involves evaluation of the reflected field at the centre of the sphere at each iteration. Although
the procedure converges and produces consistent results, for our comparative study, we would prefer to investigate a procedure that would avoid any further approximation
beyond truncation of the iterative process, so that we can be confident that the approximation works at all frequencies of oscillation. We remark however, that the analysis
of~\citet{ardekani2006Reflections} seems to be similar in content to that of section~\ref{sec:all_orders}, whereby we may expect their result to differ only in corrections
that lie beyond all orders.

Here, we consider the same geometry of a small sphere performing small oscillations near a flat wall, but with the simpler case of free-slip boundary conditions on the
wall. As before, we assume no-slip boundary conditions on the sphere. We shall truncate the iterative procedure after one reflection from the wall, but without further
approximation, yielding results that are expected to be correct to lowest order in $a/h$ for arbitrary frequency of oscillation $\omega$. The choice of full-slip boundary
conditions on the wall\footnote{Full-slip boundary conditions at solid-liquid interfaces are of increasing practical importance~\citep{vinogradova1999,neto2005}, and can be
  approximately realised on super-hydrophobic surfaces created by means of nano-fabricated structures~\citep[see e.g.][]{choi2006slip} or by increasing the surface
  roughness~\citep[see e.g.][]{shibuichi1996}. Further examples may be found in e.g.~\citet{mo2017wettability}}, as opposed to the more common no-slip / partial-slip boundary
conditions, makes the problem particularly simple as we may employ the method of images, and place an image sphere behind the wall in order to satisfy boundary conditions on
the wall. This simplicity enables exact evaluation of the surface and volume average integrals that enter the generalised Fax\'{e}n theorem (section~\ref{sec:faxen_theorem})
in closed form.

We break up the problem into two sub-problems: one with the sphere oscillating perpendicular to the wall, and the other with the sphere oscillating parallel to the wall
along any particular direction. In each case, we compute the drag force along the direction of oscillation.

In anisotropic geometries, in addition to the drag, the sphere may also experience a force in the directions normal to its motion, which would correspond to off-diagonal terms
in $\dragtensor$. We show that within the approximations used in this work, these forces are zero. In the steady case, such effects have been shown to exist when the advective
term of the Navier-Stokes equations is retained in the Oseen approximation~\citep[see e.g.][]{faxen1921thesis,shinohara1979lateral}

\subsection{Flow around a sphere oscillating in an unbounded fluid}

First, we review the well-known problem of a sphere oscillating in an unbounded fluid. The problem was first solved by~\citet{stokes1851pendulum}. However, we shall follow the
presentation of~\citet[\S 24, Prob. 5]{landaulifshitz1987fluid} as it is more convenient for our purposes\footnote{However, we shall use notation that is consistent with the
  rest of this work. This involves the changes $R \to a$, $a \to A$, $b \to B$, $-\imagunit k R \to \delta$ from the notation used in Landau-Lifshitz to our notation.}. We
had already used some of these results in section~\ref{sec:induced_force}, but the level of detail and notation here is adapted to the calculation that follows.

Using the ansatz $\vec{v}_\omega(\vec{r}) = \bnabla \times \bnabla \times (f(r) \vec{u}_\omega)$ for the velocity field $\vec{v}_\omega(\vec{r})$ generated by the sphere
oscillating with velocity $\vec{u}_\omega$, the unsteady incompressible Stokes equations~\eqref{eq:UnsteadyStokesFourier} reduce to
\begin{equation}
\Delta^2 f = \alpha^2 \Delta f,
\end{equation}
whose solution subject to the no-slip boundary conditions $\vec{v}_\omega|_{\partial S} = \vec{u}_\omega$ on the surface of the sphere and decay condition at infinity is $f(r)$ such that,
\begin{equation}
\label{eq:FDimensionful}
\frac{f'(r)}{r} = \frac{1}{r^3} \left [ A \eulere^{-\alpha r} \left (r + \frac{1}{\alpha} \right) + B \right ],
\end{equation}
with the constants,
\begin{equation}
\begin{aligned}
A &= \frac{3a^2}{2\delta} \eulere^{\delta}, \\
B &= -\frac{3 a^3}{2 \delta^2} \left ( 1 + \delta + \frac{\delta^2}{3} \right ).
\end{aligned}
\end{equation}
Here, the origin of the spherical coordinate system $(r,\theta,\varphi)$ is at the centre of the sphere, and the polar axis is along $\vec{u}_\omega$. It must be noted
that the combination $f'(r)/r$ is dimensionless. This is the same function $f'(r)/r$ from equation~\eqref{eq:FPrimeOverRFarFieldForm} written out using different notation.

From the above, the components of the velocity in the same coordinate system may be calculated as,
\begin{equation}
\begin{aligned}
\label{eq:UnboundedSphereVelocity}
\hat{e}_r \bcdot \vec{v}_{\omega} &= - 2 u_\omega \frac{f'(r)}{r} \cos \theta, \\
\hat{e}_\theta \bcdot \vec{v}_{\omega} &= u_\omega \sin \theta \left [ -\frac{A \alpha}{r} \eulere^{-\alpha r} - \frac{f'(r)}{r} \right ].
\end{aligned}
\end{equation}
It must be noted that the problem possesses axial symmetry, by which $\hat{e}_\varphi \bcdot \vec{v}_{\omega} = 0$ and there is no $\varphi$ dependence for most quantities.

\subsection{Image system for a full-slip plane wall: Perpendicular oscillations}

\begin{figure}
  \label{fig:Setup}
  \centering
  \includegraphics[width=0.75\textwidth]{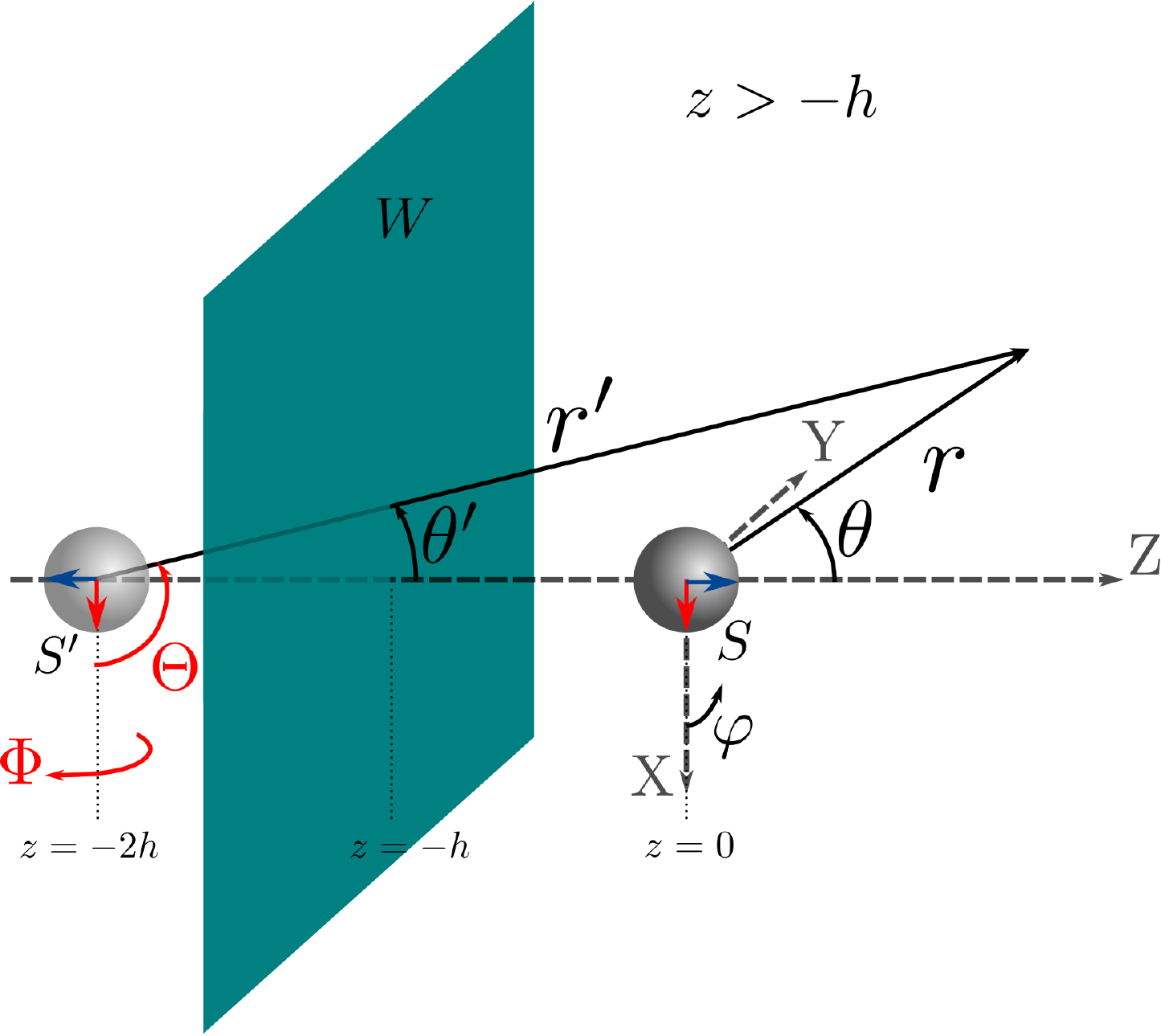}
  \caption{Image systems for oscillations perpendicular and parallel to the wall, and coordinate systems adapted to the geometry. For conciseness, we have shown both systems in a single figure. The blue horizontal arrows on the spheres, pointing in opposing directions, indicate the direction of velocities for the perpendicular case. The red vertical arrows, pointing in the same direction, indicate the same for the parallel case. The angles $\Theta$ and $\Phi$ marked in red are relevant only to the parallel case.}
\end{figure}

Let the fluid fill the half-space $\mathbb{R}^+ \times \mathbb{R}^2$ indexed by cylindrical coordinates $\rho > 0,\;z > -h,\;0 \leq \varphi < 2\thepi$ ($h>0$). Let the sphere $S$
of radius $a$ lie at the point $\rho = 0,\; z = 0$. The plane wall $W$ is located at the plane $z = -h$. For convenience, we introduce additional coordinate systems: a
spherical coordinate system $(r,\,\theta,\,\varphi)$ with origin at $z = 0$ and polar axis along the positive $z$-axis; and a spherical coordinate system
$(r',\,\theta',\,\varphi)$ with origin at $z = -2h$ and polar axis along the positive $z$-axis. Let the sphere oscillate with velocity $\vec{u}_\omega = +1 \hat{e}_z$,
where $\hat{e}_z$ is the unit vector along the positive $z$-direction. The situation is visualised in figure~\ref{fig:Setup}, where the blue horizontal arrows represent the directions of velocities. The red vertical arrows and the angles $\Theta$, $\Phi$ are irrelevant to this section.

The velocity field~\eqref{eq:UnboundedSphereVelocity} of the sphere does not satisfy the full-slip boundary conditions on the wall $W$,
\begin{equation*}
\begin{aligned}
\hat{e}_z \bcdot \vec{v}_{\omega} |_{W} &= 0, \\
\hat{e}_z \bcdot \bnabla \vec{v}_{\omega,\perp}|_{W} &= \vec{0},
\end{aligned}
\end{equation*}
where $\vec{v}_{\omega,\perp} = \vec{v}_\omega - (\hat{e}_z \bcdot \vec{v}_{\omega}) \hat{e}_z$, although it satisfies the no-slip boundary conditions
on the sphere $S$. Thus, we introduce an additional field\footnote{While there are indeed pressure fields associated with each of these velocity fields, it turns out that they
  are not directly relevant to our calculations.} $\vec{v}^{(1)}_\omega$ such that $\vec{v}_\omega + \vec{v}^{(1)}_\omega$ satisfies full-slip boundary conditions at wall $W$. The
field $\vec{v}^{(1)}_\omega$ could be regarded as the flow reflected from the wall. We could consider $\vec{v}^{(1)}_\omega$ to be produced by an \emph{image} sphere $S'$
centred at $z = -2h$ and having velocity $\vec{u'}_\omega = -1 \hat{e}_z$. By symmetry, the boundary conditions at $W$ are then satisfied. However, the combined field
$\vec{v}_\omega + \vec{v}^{(1)}_\omega$ will not satisfy the no-slip boundary conditions on $\partial S$.  Instead of computing the next reflected field $\vec{v}^{(2)}_\omega$
that corrects for the boundary conditions on the sphere, we shall simply employ $\vec{v}^{(1)}_\omega$ as the background field in the generalised Fax\'{e}n theorem~\eqref{eq:Faxen} to
calculate the drag coefficient. The iterative procedure of reflections shall be truncated at this point. Thus, it suffices to calculate the image field $\vec{v}_\omega^{(1)}$. The image
field is simply given by using~\eqref{eq:UnboundedSphereVelocity} with the replacements $u_\omega \to -1,\;\theta \to \theta',\;r \to r'$. However, in order to employ the
generalised Fax\'{e}n theorem, we would need to average this field over $\partial S$ and $S$. To do so, the following co-ordinate conversion formulas are handy,
\begin{equation}
\begin{aligned}
\label{eq:CoordConversionPerp}
  \rho &= r' \sin \theta' = r \sin \theta, \\
  z &= r \cos \theta = r' \cos \theta' - 2h, \\
  r'^2 &= 4 h^2 + r^2 + 4 r h \cos \theta, \\
  r^2 &= \rho^2 + z^2, \\
  r'^2 &= \rho^2 + (2h + z)^2.
\end{aligned}
\end{equation}
It is also convenient to introduce the non-dimensionalized variables, $\xi := r'/(2h)$, $\delta = \alpha a$, $\epsilon = a/h$. Then we may write the dimensionless function
$F_0(\xi) := f'(r')/r'$ , i.e. the function of equation~\eqref{eq:FDimensionful} evaluated instead at $r'$, as
\begin{equation}
F_0(\xi) = \frac{1}{\xi^3} \left [ \constp \, \eulere^{-2 \delta \xi/\epsilon} \left( 1 + \frac{2\delta \xi}{\epsilon} \right) - \constq \right ],
\end{equation}
where the constants $\constp := 3 \epsilon^3 \eulere^\delta / (16 \delta^2)$ and $\constq := 3 \epsilon^3 (1 + \delta + \delta^2 / 3) / (16 \delta^2)$.

\subsection{Drag coefficient for perpendicular oscillations}

With these preparations, we are ready to calculate the drag force on the sphere for oscillations perpendicular to the wall. To do so, we need to compute the averages of the
first reflected field $\vec{v}^{(1)}_\omega = -\bnabla \times \bnabla \left [ f(r') \hat{e}_z \right ]$ on $\partial S$ and $S$. For this purpose, it is convenient to
leave $\vec{v}^{(1)}_\omega$ in this form rather than expand it out as in equation~\eqref{eq:UnboundedSphereVelocity}. By symmetry, we observe that the only non-vanishing
contribution comes from the $z$-component
\begin{equation}
V := \hat{e}_z \bcdot \vec{v}^{(1)}_{\omega} = - \hat{e}_z \bcdot \bnabla \left ( \hat{e}_z \bcdot \bnabla f(r') \right ) + \Delta f(r').
\end{equation}
We begin by computing the average over the surface of a sphere of radius $r = a \zeta$ centred about $z = 0$, given by,
\begin{equation}
\bar{V}^S(\zeta) = \frac{1}{4 \thepi} \int_{0}^{\thepi} 2 \thepi \sin \theta \,\mathrm{d}\theta \; V,
\end{equation}
where we have already performed the trivial $\mathrm{d}\varphi$ integral.

Writing $\bnabla$ in the cylindrical coordinate system\footnote{We will frequently ignore the $\varphi$ derivatives in these expressions as they are zero due to axial
  symmetry.} as
\begin{equation*}
\bnabla = \vec{\hat{e}}_z \left ( \frac{\partial}{\partial z} \right )_\rho + \vec{\hat{e}}_\rho \left ( \frac{\partial}{\partial \rho} \right )_z,
\end{equation*}
we find from~\eqref{eq:CoordConversionPerp} that $\left ( {\partial r'}/{\partial z} \right )_\rho = (z + 2h)/r'$, and use this
in the expression for $V$ to obtain,
\begin{equation}
\begin{aligned}
V &= - \vec{\hat{e}}_z \bcdot \bnabla \left [ \frac{z + 2h}{r'} f'(r') \right ] + f''(r') + 2\frac{f'(r')}{r'} \\
&= - \frac{(z + 2h)^2}{r'}\der{r'}\left [ \frac{f'(r')}{r'} \right ] + f''(r') + \frac{f'(r')}{r'} \\
& = -\frac{(z + 2h)^2}{r'}\der{r'}\left[ \frac{f'(r')}{r'} \right ] + \frac{1}{r'}\der{r'}\left[ r' f'(r') \right].
\end{aligned}
\end{equation}

We now observe from~\eqref{eq:CoordConversionPerp} that since we are integrating on a surface of constant $r$, $\frac{1}{2} \sin \theta \, \mathrm{d}\theta = - (2 r)^{-1}\, \mathrm{d}z = - (\epsilon \zeta)^{-1}
\xi\,\mathrm{d}\xi$, whereby the integral may be rewritten in terms of the non-dimensionalized variables as,
\begin{equation}
\begin{aligned}
\bar{V}^S(\zeta) &= \frac{1}{\epsilon \zeta} \int_{1 - \frac{1}{2}\epsilon \zeta}^{1 + \frac{1}{2}\epsilon\zeta} \xi\, \mathrm{d}\xi\; V \\
&= \frac{1}{\epsilon \zeta} \int_{1 - \frac{1}{2}\epsilon\zeta}^{1 + \frac{1}{2}\epsilon\zeta} \mathrm{d}\xi \, \left \{ -\frac{1}{4} \left ( \xi^2 - \frac{\epsilon^2
      \zeta^2}{4} + 1 \right )^2\der{\xi}\,[F_0(\xi)] + \der{\xi}\left[ \xi^2 F_0(\xi) \right] \right \}.
\end{aligned}
\end{equation}
The advantage of this form is that the integral may be conveniently evaluated using integration by parts, and with the definitions,
\begin{equation}
\begin{aligned}
F_1(\xi) &:= \int \xi F_0(\xi)\,\mathrm{d}\xi = -\frac{1}{\xi} \left [ \constp\, \eulere^{-2 \delta \xi / \epsilon} - \constq \right ], \\
F_2(\xi) &:= \int \xi F_1(\xi)\,\mathrm{d}\xi = \left [ \constp \frac{\epsilon}{2\delta}\eulere^{-2 \delta \xi / \epsilon} + \constq \xi \right ],
\end{aligned}
\end{equation}
we have,
\begin{equation}
\begin{aligned}
\bar{V}^S(\zeta) &= -\frac{2}{\epsilon \zeta} \left [ F_2(\xi) - \xi F_1(\xi) \right ]^{1 + \frac{1}{2}\epsilon\zeta}_{1 - \frac{1}{2}\epsilon\zeta} \\
&= \frac{2}{\epsilon\zeta} \left [ 2 \constp \, \eulere^{-2\delta/\epsilon}\sinh ( \delta \zeta ) \left ( 1 + \frac{\epsilon}{2\delta} \right ) - \constq \epsilon \zeta \right ].
\end{aligned}
\end{equation}
The average $\bar{V}^S$ on the surface of the sphere $\partial S$ is just obtained by evaluating the above at $\zeta = 1$.

We define the volume average of V,
\begin{equation}
\begin{aligned}
\bar{V}^V &:= \frac{1}{\frac{4}{3} \thepi a^3} \int_0^a 4 \thepi r^2 \,\mathrm{d}r\: \bar{V}^S(r/a) \\
&= \int_0^1 3 \zeta^2\,\mathrm{d}\zeta \: \bar{V}^S(\zeta),
\end{aligned}
\end{equation}
which may be evaluated to obtain,
\begin{equation}
  \bar{V}^V = \frac{12 \constp}{\epsilon \delta^2} \left ( 1 + \frac{\epsilon}{2\delta} \right ) \eulere^{-2\delta / \epsilon} \left (\delta \cosh \delta - \sinh \delta \right ) - 2
  \constq.
\end{equation}
We now rewrite the generalised Fax\'{e}n theorem~\eqref{eq:Faxen} as,
\begin{equation}
\label{eq:FaxenForDragCoefficient}
\frac{\gamma_{\perp}^{R}}{\gamma_s} = \frac{\gamma_0}{\gamma_s} - \left [ (1 + \delta) \bar{V}^S + \frac{\delta^2}{3} \bar{V}^V \right ],
\end{equation}
where we have introduced the superscript $R$ to distinguish the results from the method of reflections from the other methods considered in this work. We then use the above to
obtain the drag coefficient $\gamma^R_\perp$ as,
\begin{equation}
\label{eq:GammaPerpendicularReflections}
  \frac{\gamma_\perp^R}{\gamma_s} = \left ( 1 + \delta + \frac{\delta^2}{9} \right ) + \frac{3 \epsilon}{8 \delta^2} \left [ \epsilon^2 \left ( 1 + \delta + \frac{\delta^2}{3}
    \right )^2 - \eulere^{2\delta ( 1 - 1/\epsilon)} ( 2 \epsilon \delta + \epsilon^2 ) \right ].
\end{equation}

\subsection{Image system for a full-slip plane wall: Parallel oscillations}

As before, we consider the fluid to fill the half-space $\mathbb{R}^+ \times \mathbb{R}^2$. We will instead prefer to use a Cartesian coordinate system $(x,\,y,\,z)$ where the
half-space occupied by the fluid corresponds to $z > -h$ ($h > 0$). Let the sphere $S$ of radius $a$ lie at the origin of the Cartesian coordinate system. The plane wall $W$
is located at $z = -h$. For convenience, as before, we introduce additional coordinate systems: a spherical coordinate system $(r,\,\theta,\,\varphi)$ with origin at $z = 0$
and polar axis along the positive $z$-axis; a spherical coordinate system $(r',\,\theta',\,\varphi)$ with origin at $z = -2h$ and polar axis along the positive $z$-axis; and
another spherical coordinate system $(R := r',\,\Theta,\,\Phi)$ with origin at $z = -2h$ and polar axis along the positive $x$-axis.  Let the sphere oscillate with velocity
$\vec{u}_\omega = +1 \vec{\hat{e}}_x$, where $\hat{e}_x$ is the unit vector along the $x$-direction. The situation is visualised in figure~\ref{fig:Setup}, where the red vertical arrows represent the directions of velocities. The blue horizontal arrows are irrelevant to this section.

As before, we introduce an image sphere $S'$ centred at $z = -2h$, but to satisfy the boundary conditions on $z = -h$, the image sphere must have the same velocity as the
actual sphere, i.e. $\vec{u'}_\omega = +1 \vec{\hat{e}}_x$. We list the relevant coordinate conversion formulas involving the $(x,\,y,\,z)$ and the $(R,\,\Theta,\,\Phi)$
systems below:
\begin{equation}
\begin{aligned}
\label{eq:ParallelCoordConversionPerp}
R &= r' \\
x &= r' \sin \theta' \sin \varphi.
\end{aligned}
\end{equation}

\subsection{Drag coefficient for parallel oscillations}

We now proceed to calculate the drag force on the sphere for oscillations parallel to the wall. The first reflected field is now given by,
\begin{equation}
\vec{v}^{(1)}_\omega = +\bnabla \times \bnabla \left [ f(R) \vec{\hat{e}}_x \right ].
\end{equation}

The relevant component is the $x$-component,
\begin{equation}
V := \hat{e}_x \bcdot \vec{v}^{(1)}_{\omega} = + \vec{\hat{e}}_x \bcdot \bnabla \left ( \vec{\hat{e}}_x \bcdot \bnabla f(R) \right ) - \Delta f(R).
\end{equation}
While there is no immediate reason to preclude the drag force from having a $z$-component, we will later show that there is none in the first-reflection approximation that we
compute here.

The average over the surface of a sphere of radius $r = a \zeta$ centred about $z = 0$ is given by,
\begin{equation}
\label{eq:VSDefineParallel}
\bar{V}^S(\zeta) = \frac{1}{4 \thepi} \int_{\theta = 0}^{\thepi} \int_{\varphi = 0}^{2\thepi} \sin \theta \,\mathrm{d}\theta \; \mathrm{d}\varphi \; V,
\end{equation}
as we do not have azimuthal symmetry in this case.

Writing $\bnabla$ in the Cartesian coordinate system as
\begin{equation*}
\bnabla = \vec{\hat{e}}_x \left ( \frac{\partial}{\partial x} \right )_{y,\,z} + \vec{\hat{e}}_x \left ( \frac{\partial}{\partial y} \right )_{x,\,z} + \vec{\hat{e}}_z \left (
  \frac{\partial}{\partial z} \right )_{x,\,y},
\end{equation*}
we find from the coordinate conversion formulas~\eqref{eq:CoordConversionPerp} that $( {\partial r'}/{\partial x} )_{y,z} = {x}/{r'}$, and use this in the
expression for $V$ to obtain,
\begin{equation}
\begin{aligned}
\label{eq:VParallel}
V &= \vec{\hat{e}}_x \bcdot \bnabla \left [ \frac{x}{r'} f'(r') \right ] - f''(r') - \frac{2 f'(r')}{r'} \\
  &= \frac{x^2}{r'} \der{r'} \left [ \frac{f'(r')}{r'} \right ] - f''(r') - \frac{f'(r')}{r'} \\
  &= \frac{x^2}{r'} \der{r'} \left [ \frac{f'(r')}{r'} \right ] - \frac{1}{r'} \der{r'} \left [ r' f'(r') \right ].
\end{aligned}
\end{equation}

We now write $x = r \sin \theta \cos \varphi$ in the expression for V and observe that $\int_0^{2\thepi} \mathrm{d}\varphi \; \cos^2 \varphi = \thepi$, whereby we may reduce
\eqref{eq:VSDefineParallel} to,
\begin{equation}
\bar{V}^S(\zeta) = \frac{r^2}{4} \int_{0}^{\thepi} \mathrm{d}\theta \; \sin^3\theta \; \frac{1}{r'} \der{r'} \left [ \frac{f'(r')}{r'} \right ] -
\frac{1}{\epsilon \zeta} \int_{1-\epsilon \zeta/2}^{1+\epsilon \zeta/2} \mathrm{d}\xi \; \der{\xi} \left [\xi^2 F_0(\xi) \right ],
\end{equation}
where we have treated the second term in equation~\eqref{eq:VParallel} as we did in the case of perpendicular oscillations. For the first integral in
the above equation, we note that since $r$ is constant, we may write
$r^2 \sin^3 \theta \, \mathrm{d}\theta = (r \sin \theta \, \mathrm{d}\theta) r (1 - \cos^2 \theta) = - \mathrm{d}z \, (r^2 - z^2)/r$ and substitute for $z$ in terms of $r'$ to
obtain,
\begin{equation}
\begin{aligned}
\frac{r^2}{4} \int_{0}^{\thepi} \mathrm{d}\theta \; \sin^3 \theta \; \frac{1}{r'} \der{r'} \left [ \frac{f'(r')}{r'} \right ] &= \frac{1}{4r}
\int_{1-\epsilon \zeta/2}^{1+\epsilon \zeta/2} \frac{\mathrm{d}r'}{2h}\, r'^2 \der{r'} \left [ \frac{f'(r')}{r'} \right ] \\
&+ \frac{1}{2} \left [ - \frac{1}{2r} \int_{-r}^{r} \frac{\mathrm{d}z}{r'} (2h + z)^2 \der{r'} \left [ \frac{f'(r')}{r'} \right ] \right ].
\end{aligned}
\end{equation}
The second integral in the above expression was previously evaluated for the perpendicular case, so we may simply use the result. In non-dimensionalized variables, the first
integral has the form
\begin{equation}
\frac{1}{2\epsilon \zeta} \int_{1-\epsilon\zeta/2}^{1+\epsilon\zeta/2} \mathrm{d}\xi \, \xi^2 \der{\xi} F_0(\xi),
\end{equation}
which may be easily integrated by parts and expressed in terms of $F_1(\xi)$. Thus, we have
\begin{equation}
  \bar{V}^S(\zeta) = \frac{1}{\epsilon \zeta} \left [ (\xi - 1) F_1(\xi) - F_2(\xi) - \xi^2 F_0(\xi) \right ]_{1-\epsilon\zeta/2}^{1+\epsilon\zeta/2},
\end{equation}
which simplifies to,
\begin{equation}
  \bar{V}^S(\zeta) = 2 \constp \, \eulere^{-2\delta/\epsilon} \frac{\sinh (\delta \zeta )}{\epsilon \zeta} \left ( 1 + \frac{2 \delta}{\epsilon} +
    \frac{\epsilon}{2\delta} \right ) - \constq.
\end{equation}
The average $\bar{V}^S$ on the surface of the sphere $\partial S$ is just obtained by evaluating the above at $\zeta = 1$.

As before, the volume average of V may be obtained,
\begin{equation}
\bar{V}^V = \frac{6 \constp}{\epsilon \delta^2} \left ( 1 + \frac{2 \delta}{\epsilon} + \frac{\epsilon}{2\delta} \right ) \eulere^{-2\delta/\epsilon} \left [ \delta \cosh \delta
 - \sinh \delta \right ] - \constq.
\end{equation}

We now adapt the generalised Fax\'{e}n theorem~\eqref{eq:Faxen} as we did in equation~\eqref{eq:FaxenForDragCoefficient} to obtain the drag coefficient $\gamma_{\|}^R$ as,
\begin{equation}
\label{eq:GammaParallelReflections}
\frac{\gamma_{\|}^R}{\gamma_s} = \left (1 + \delta + \frac{\delta^2}{9} \right ) + \frac{3 \epsilon}{
  16 \delta^2} \left [  \epsilon^2 \left (1 + \delta + \frac{\delta^2}{3} \right )^2 -
    \eulere^{2 \delta (1 - 1/\epsilon)} (4 \delta^2 +
       2 \epsilon \delta + \epsilon^2 ) \right ].
\end{equation}

We will now show that there is no force along the $z$-direction to first order. The $z$-component of the first reflected field due to parallel oscillations of the sphere
is given by,
\begin{equation}
\begin{aligned}
v^{(1)}_{xz,\,\omega} &= \vec{\hat{e}}_z \bcdot \bnabla \left ( \vec{\hat{e}}_x \bcdot \bnabla f(r') \right ) - \vec{\hat{e}}_x \bcdot \vec{\hat{e}}_z \Delta f(r') \\
&= \vec{\hat{e}}_z \bcdot \bnabla \left ( \frac{x}{r'} f'(r') \right ) - 0 \\
&= \frac{(z+2h)x}{r'} \der{r'} \left[ \frac{f'(r')}{r'} \right ].
w\end{aligned}
\end{equation}
Substituting $x = r\sin \theta \cos \varphi$ as before, we see that the surface average $\bar{V}^S$ contains the integral $\int_0^{2\thepi} \mathrm{d}\varphi \, \cos \varphi =
0$. Thus, the surface average vanishes on any spherical surface centred about $z = 0$, and consequently, the volume integral over the sphere $S$ also vanishes.

\begin{figure}
\centering
\includegraphics[width=\textwidth]{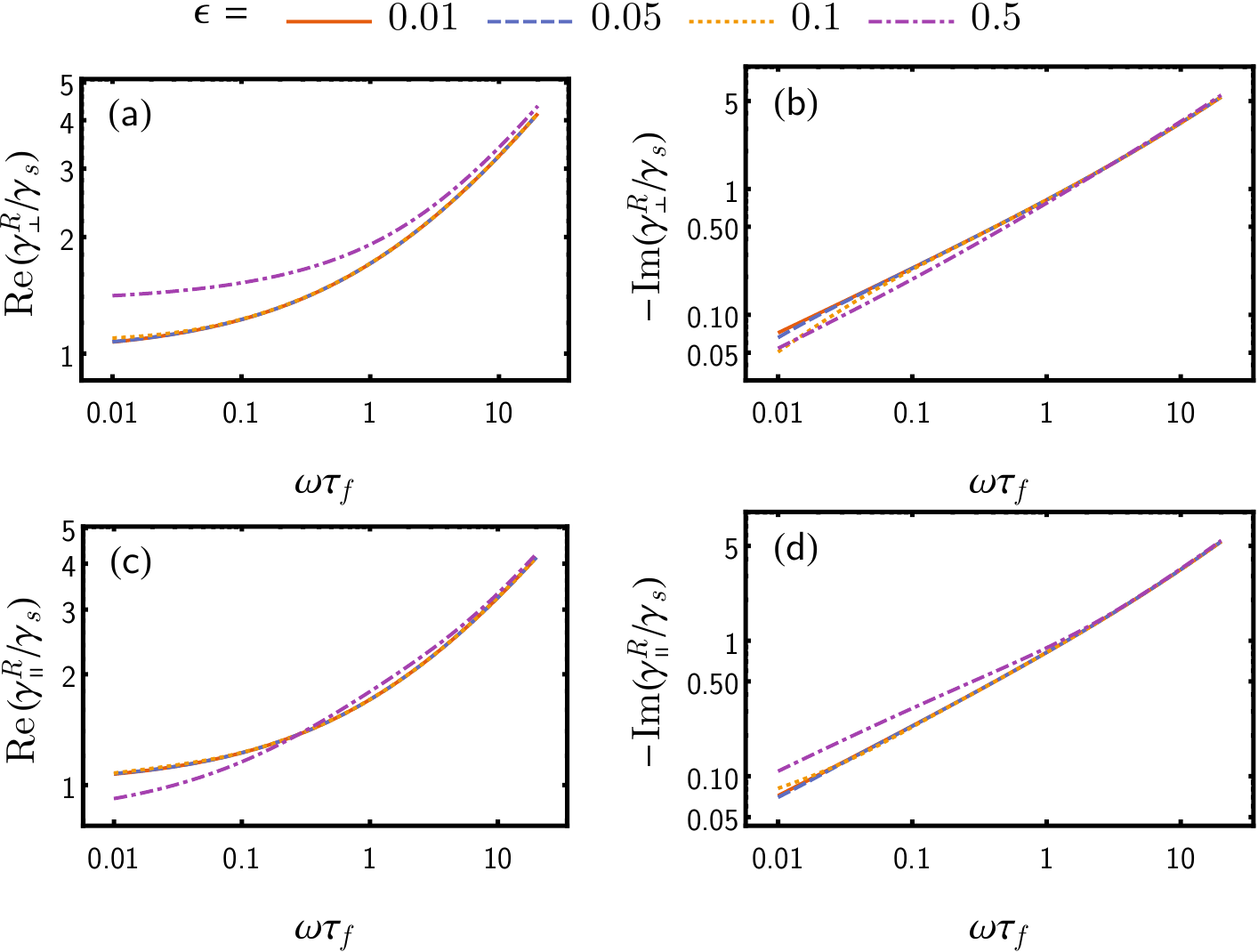}
\caption{\label{fig:GammaReflections}Logarithmic plots of the normalised drag coefficient for a no-slip sphere (radius $a$) in a viscous fluid near a full-slip plane wall (distance $h$), for various
  values of $\epsilon = a/h$, obtained using the method of reflections (\S \ref{sec:reflection_method}) in the perpendicular direction~\eqref{eq:GammaPerpendicularReflections} (a)
  real part and (b) negative imaginary part, and in the parallel direction~\eqref{eq:GammaParallelReflections} (c) real part and (d) negative imaginary part. The drag coefficient is normalised
  to the steady free-space Stokes drag coefficient $\gamma_s$. The horizontal axis is the non-dimensionalized frequency of oscillation of the sphere
  $\omega \tau_f$, where $\tau_f = a^2 \rho_f / \eta$ is the timescale for vorticity diffusion over the size of the sphere.}
\end{figure}

The results from equations~\eqref{eq:GammaPerpendicularReflections} and~\eqref{eq:GammaParallelReflections} are plotted in figure~\ref{fig:GammaReflections} as a function of
the non-dimensionalized frequency $\omega \tau_f = \imagunit \delta^2$, where $\tau_f := a^2 \rho_f / \eta$ is the time-scale over which vorticity diffuses over the size of
the sphere~\citep{franosch2011resonances}.

\section{Comparison of the point-particle approximation and the method of reflections}
\label{sec:comparison_of_results}

In this section, we compare results for the drag coefficient for a sphere near a full-slip plane wall obtained by the two methods considered earlier, viz. the point-particle
approximation (\S\ref{sec:point_particle_approximation}) and the method of reflections (\S\ref{sec:reflection_method}). Where relevant, we will also compare our
modified point-particle approximation with the point-particle approximation as used by~\citet{felderhof2012PhysRevE}.

While we may directly use the expressions for the reaction field tensor from~\citet{felderhof2012PhysRevE} in
equation~\eqref{eq:DragCoefficientTensorFromReactionFieldTensor} to compute the drag coefficients in the parallel and perpendicular directions, it is however useful for
purposes of comparison to first put the expression for the drag coefficient in a form similar to those obtained using the method of reflections in
equations~\eqref{eq:GammaPerpendicularReflections} and~\eqref{eq:GammaParallelReflections}. To effect this, we first assume that $\gamma_s \tens{R}$ is small (which we would
expect to be true on physical grounds in the regime of validity of the point-particle approximation), whereby we may
expand~\eqref{eq:DragCoefficientTensorFromReactionFieldTensor} to first order in $\gamma_s \tens{R}$ to obtain\footnote{We remark that this form is likely inferior for
  numerical computations, since the original expression was in the form of a Pad\'{e} approximant, which has been observed in many cases to perform better (see
  section~\ref{sec:numerical_comparison}).},
\begin{equation}
\begin{aligned}
\label{eq:DragCoefficientTensorFromReactionFieldTensorApproximate}
\pmb{\gamma} &= \gamma_0(\omega) \left [ \mathds{1} - ( 1 + \delta + \delta^2 / 3 ) \gamma_s \tens{R} \right ] - \frac{2}{9} \gamma_s \delta^2 (1 + \delta + \delta^2 / 3 )
\gamma_s \tens{R} + o[\gamma_s \tens{R}]\\
&= \gamma_0(\omega) \mathds{1} - \gamma_s ( 1 + \delta + \delta^2/3 )^2 (\gamma_s \tens{R}) + o[\gamma_s \tens{R}].
\end{aligned}
\end{equation}

Plugging in the expressions from~\citet[eq. (3.5) and (3.16)]{felderhof2012PhysRevE},
\begin{equation}
\label{eq:FelderhofReactionField}
\begin{aligned}
\gamma_s R_{zz} &= \frac{3 \epsilon}{2} \left \{ -\frac{1}{4\nu^2} \left [ 1 - (1 + 2\nu) \eulere^{-2\nu} \right ] \right \}, \\
\gamma_s R_{xx} &= \frac{3 \epsilon}{2} \left \{ -\frac{1}{8\nu^2} \left [ 1 - (1 + 2\nu + 4\nu^2) \eulere^{-2\nu} \right ] \right \},
\end{aligned}
\end{equation}
for the components $R_{zz} := \hat{e}_z \bcdot \tens{R} \bcdot \hat{e}_z$ and $R_{xx} := \hat{e}_x \bcdot \tens{R} \bcdot \hat{e}_x$ of $\tens{R}$, where $\nu := \alpha h = \delta / \epsilon$, into the above expression, we obtain the expressions
\begin{equation}
\begin{aligned}
\label{eq:PointParticleDragTensorApproximated}
\frac{\gamma_{\perp}^P}{\gamma_s} &\approx \left ( 1 + \delta + \frac{\delta^2}{9} \right ) + \frac{3 \epsilon}{8 \delta^2} \left ( 1 + \delta + \frac{\delta^2}{3} \right )^2\;
\left [ \epsilon^2 - \eulere^{-2\delta/\epsilon} \left ( 2 \epsilon \delta + \epsilon^2 \right ) \right ] \\
\frac{\gamma_\|^P}{\gamma_s} &\approx \left ( 1 + \delta + \frac{\delta^2}{9} \right ) + \frac{3 \epsilon}{16 \delta^2} \left ( 1 + \delta + \frac{\delta^2}{3} \right )^2\; \left
  [ \epsilon^2 - \eulere^{-2\delta/\epsilon} \left ( 4 \delta^2 + 2 \epsilon \delta + \epsilon^2 \right ) \right ].
\end{aligned}
\end{equation}

We now compare these against equations~\eqref{eq:GammaPerpendicularReflections} and~\eqref{eq:GammaParallelReflections} to find that the expressions from the two methods
indeed differ, but in the factor in front of the subdominant (as $\epsilon \to 0^+$, $\delta$ fixed) exponential term $\eulere^{-2\delta/\epsilon}$. We shall show in the following sub-sections that
in the regimes where the exponential terms actually matter, the two results agree to first order in $\epsilon$. Thus, unless the physics under investigation expressly relies
on the subdominant terms, the results from the two methods agree to first order.

\subsection{Asymptotic comparison}
\label{sec:asymptotic_comparison}

Since there are two length scales, there are four asymptotic regimes that we may consider, depending on how $\alpha$ compares with $a$ and $h$. Of particular interest here are
two regimes -- the regime of low frequencies where $\alpha h \sim 1$, and that of high frequencies where $\alpha a \gg 1$. The former regime is of interest
owing to our discussion about the subdominant exponential terms (sections~\ref{sec:all_orders} and~\ref{sec:comparison_of_results}). The latter regime is of interest owing to
the discrepancy in effective mass mentioned in the introduction. It can be easily verified that the results from the method of reflections as well as the modified
point-particle approximation agree in the regime of intermediate frequencies $\alpha a \sim 1$.

\subsubsection{Low frequencies}
We now consider non-zero, but low frequencies, where $\nu := \alpha h \sim 1$ but $\delta = \alpha a \ll 1$, i.e. the skin-depth of vorticity is comparable to the sphere-wall
separation, and is much larger than the size of the sphere.

In the point-particle calculation, no approximation can be made in the expressions for the reaction field tensor~\citep[equations (3.5) and (3.16)]{felderhof2012PhysRevE} in
this regime. However, we substitute $\delta = \epsilon \nu$ in~\eqref{eq:DragCoefficientTensorFromReactionFieldTensor} and keep terms to first order in $\epsilon$ while noting
that $\gamma_s \tens{R}$ is first order in $\epsilon$ to obtain,
\begin{equation}
\dragtensornoarg \sim  \gamma_s \left [ \mathds{1}( 1 + \epsilon \nu)  - \gamma_s \tens{R} \right ],
\end{equation}
which, upon substitution for the components of $\tens{R}$ yields
\begin{equation}
\begin{aligned}
\frac{\gamma_\perp^P}{\gamma_s} &\sim 1 + \epsilon \nu + \frac{3 \epsilon}{8\nu^2} \left [ 1 - \left ( 1 + 2\nu \right ) \eulere^{-2\nu} \right ] + o(\epsilon), \\
\frac{\gamma_\|^P}{\gamma_s} &\sim 1 + \epsilon \nu + \frac{3 \epsilon}{16 \nu^2} \left [ 1 - \left ( 1 + 2\nu + 4\nu^2 \right ) \eulere^{-2\nu} \right ] + o(\epsilon).
\end{aligned}
\end{equation}

For the results from the method of reflections, we once again substitute $\delta = \epsilon \nu$ in equations~\eqref{eq:GammaPerpendicularReflections}
and~\eqref{eq:GammaParallelReflections} and keep terms to first order in $\epsilon$, and obtain the same results as above for $\gamma_\perp^R$ and $\gamma_\|^R$.

Thus, even where the subdominant exponential terms are important, the two results agree to lowest order in $\epsilon$.

We may also take the $\alpha \to 0$ limit in the above and as expected, we recover expressions that agree with results obtained through image systems for steady Stokes
flows~\citep{frydel2006latticeboltzmann}.

\subsubsection{High frequencies}
\label{sec:potential_flow}

We finally consider the range of frequencies $\omega \gg \eta / (\rho_f a^2)$, where $1/|\alpha| \ll a \ll h$. In this regime, we expect that the viscous contributions to the
drag coefficient are negligible compared to the inertial contributions, i.e. the added mass term. For instance, in the case of a spherical particle in an unbounded fluid medium, the drag
coefficient in this regime $\gamma_0(\omega) \sim \gamma_s \delta^2 / 9 = - \imagunit \omega m_f / 2$, which is the added mass contribution from the fluid.

The added mass of a particle executing small oscillations in a fluid is usually obtained by means of potential flow~\citep[see
e.g.][]{landaulifshitz1987fluid,brennen1982review}. In particular, the added mass of a spherical particle near a plane wall is a well-studied
problem~\citep{lamb1932hydrodynamics,milne1968theoretical,yang2010addedmass}, and the expressions for the effective masses in this case,
\begin{equation}
\begin{aligned}
\label{eq:EffectiveMassPlaneWall}
m^\ast_\perp &= m_p + \frac{m_f}{2} \left [ 1 + \frac{3}{8} \left (\frac{a}{h} \right )^3 \right ], \\
m^\ast_{\|} &= m_p + \frac{m_f}{2} \left [ 1 + \frac{3}{16} \left (\frac{a}{h} \right )^3 \right ],
\end{aligned}
\end{equation}
are well-known. It must be noted that owing to the absence of $\Delta \vec{v}_\omega$, the differential equation is of lower order, whereby fewer boundary conditions are
needed for the potential flow calculation, and thus the added mass obtained from potential flow does not distinguish between full-slip and no-slip boundary conditions.

It has been pointed out~\citep{mo2015pre} that Felderhof's expressions do not agree with these results. As pointed out by~\citet{zwanzig1975}, the velocity
auto-correlation function for a Brownian particle in an incompressible fluid asymptotes to $k_B T / m^\ast$ as $t \to 0$, where $m^\ast$ is the effective mass of the particle
in the fluid.\footnote{The apparent contradiction with the energy equipartition theorem, which reports a $k_B T / m$ asymptote, is resolved by including the effects of
  compressibility.}  This has been verified in unbounded fluid by experiments~\citep{kheifets2014science,mo2015opex}. However, the results from~\citet[eq. (4.5)]{felderhof2005JPhysChemB} (see also erratum~\citet{felderhof2005erratum}) suggest values for the added masses as
$(m_f/2)( 1 + a^3/(8h^3)  + o(a^3/h^3) )$ and $(m_f/2) ( 1 + a^3/(16h^3) + o(a^3/h^3))$. As we will presently demonstrate, this discrepancy is resolved by our modification of the point-particle framework
described in sections~\ref{sec:induced_force} and~\ref{sec:consequences_correction}.

In our modified point-particle framework, we take the asymptotics of the components of the reaction field tensor as $\nu \to \infty$ to obtain
\begin{equation}
\begin{aligned}
\gamma_s R_{zz} &\sim -\frac{3 \epsilon}{8 \nu^2}, \\
\gamma_s R_{xx} &\sim -\frac{3 \epsilon}{16 \nu^2}.
\end{aligned}
\qquad (\nu \to +\infty)
\end{equation}
We then replace $\delta = \epsilon \nu$ in~\eqref{eq:DragCoefficientTensorFromReactionFieldTensor}, substitute the above asymptotic forms for the components of $\tens{R}$, and
expand to lowest order in $\epsilon$ to obtain,
\begin{equation}
\begin{aligned}
\frac{\gamma_\perp^P}{\gamma_s} &\sim \frac{1}{9} \epsilon^2 \nu^2 \left ( 1 + \frac{3}{8} \epsilon^3 \right ) + o(\epsilon^5),\\
\frac{\gamma_\|^P}{\gamma_s} &\sim \frac{1}{9} \epsilon^2 \nu^2 \left ( 1 + \frac{3}{16} \epsilon^3 \right ) + o(\epsilon^5).
\end{aligned}
\qquad (\nu \to +\infty)
\end{equation}
Thereafter, identifying $\epsilon^2 \nu^2 / 9 = -\imagunit \omega m_f / 2$, we obtain added masses consistent with the effective masses given in~\eqref{eq:EffectiveMassPlaneWall}.

For the results from the method of reflections, we take the asymptotic as $\delta \to \infty$ with fixed $\epsilon$ in equations~\eqref{eq:GammaPerpendicularReflections}
and~\eqref{eq:GammaParallelReflections}. The subdominant exponential terms drop and we are left with
\begin{equation}
\begin{aligned}
\frac{\gamma_\perp^R}{\gamma_s} &\sim \frac{1}{9} \delta^2 \left ( 1 + \frac{3}{8} \epsilon^3 \right ), \\
\frac{\gamma_\|^R}{\gamma_s} &\sim \frac{1}{9} \delta^2 \left ( 1 + \frac{3}{16} \epsilon^3 \right ),
\end{aligned}
\qquad (\delta \to +\infty)
\end{equation}
which are once again consistent with the results from the modified point-particle approximation and with calculations from potential flow~\citep{lamb1932hydrodynamics,milne1968theoretical,brennen1982review}.

\subsection{Numerical comparison}
\label{sec:numerical_comparison}

In this section, we present numerical comparisons of predictions for the drag coefficients from three methods -- the point-particle approximation proposed
by~\citet{felderhof2012PhysRevE}, the modified point-particle approximation presented in this work (\S~\ref{sec:point_particle_approximation}),
and the method of reflections (\S~\ref{sec:reflection_method}).

Generally speaking, for purposes of numerical evaluation, it is likely that keeping the expression for $\dragtensor$ in the form of a Pad\'{e} approximant as in equation
\eqref{eq:DragCoefficientTensorFromReactionFieldTensor} gives better results. In the context of the method of reflections for the steady Stokes
equations,~\citet[][chap. 7]{happel1965} suggest the use of a geometric series extrapolation to account for higher order reflections in the absence of any further information,
which essentially amounts to turning the result from the method of reflections into a Pad\'{e} approximant. We also noted this when we computed the perturbative result to all
orders in section~\ref{sec:all_orders}. Several experiments have employed the Pad\'{e} form of the steady drag~\citep[][Fig. 2]{schaffer2007surface,mo2015pre} with good
results.

\begin{figure}
\centering
\includegraphics[width=\textwidth]{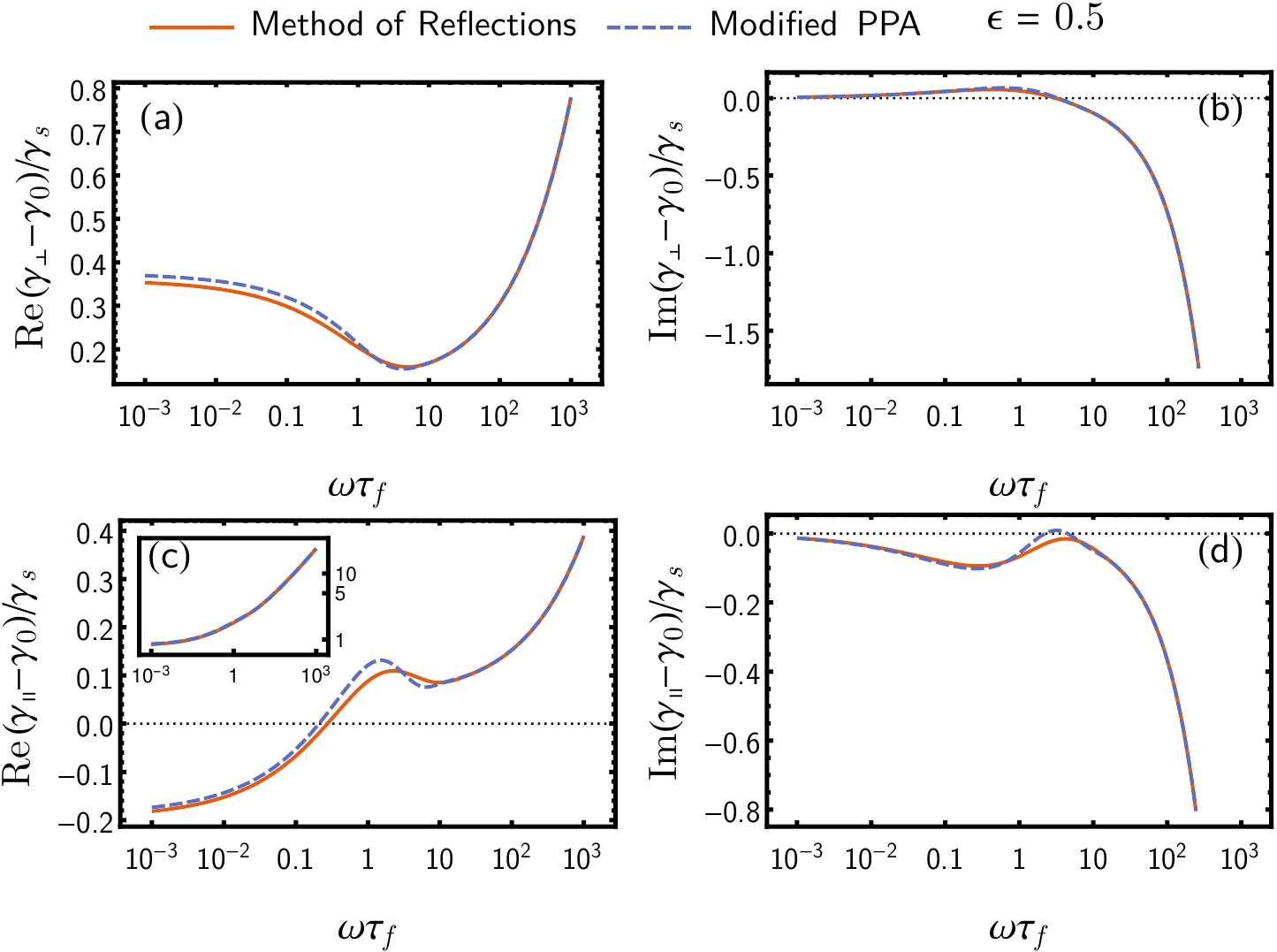}
\caption{\label{fig:ReflectionsVsPPA}Semi-logarithmic plots comparing the results for the drag coefficient of a no-slip sphere near a full-slip plane wall ($\epsilon = 0.5$)
  from the method of reflections (\S \ref{sec:reflection_method}) and the modified point-particle approximation (``Modified PPA'') (\S \ref{sec:point_particle_approximation})
  in the perpendicular direction (a) real part and (b) imaginary part, and in the parallel direction (c) real part and (d) imaginary part. In each case, the free-space drag
  coefficient $\gamma_0(\omega)$ has been subtracted in order to clearly highlight the small differences, and the coefficients have been normalised by $\gamma_s$. The inset in (c)
  shows a log-log plot of the real parts of the normalised drag coefficients in the parallel direction without subtraction of $\gamma_0$.}
\end{figure}

However, in order to appropriately compare and highlight the differences between the theories, it is necessary that we compare results expressed in similar forms. In the plots
that follow, when comparing the method of reflections against the modified point-particle approximation (figure~\ref{fig:ReflectionsVsPPA}
\suppress{and figure~\ref{fig:ReflectionsVsPPAError}}), we use the form of~\eqref{eq:PointParticleDragTensorApproximated} for the point-particle approximation. When comparing the modified
point-particle approximation against that of~\cite{felderhof2012PhysRevE} (figures~\ref{fig:PPAVsFelderhof} and~\ref{fig:PPAVsFelderhofError}), we shall use the
original forms from equations~\eqref{eq:DragCoefficientTensorFromAdmittanceUnmodified} and~\eqref{eq:DragCoefficientTensorFromReactionFieldTensor}.

We obtain the drag coefficients from Felderhof's point-particle framework by setting the expression for the admittance from~\citet[][equation
(2.9)]{felderhof2012PhysRevE} equal to~\eqref{eq:AdmittanceWithoutTrap}:
\begin{equation}
\label{eq:DragCoefficientTensorFromAdmittanceUnmodified}
\dragtensornoarg^F(\omega) = \imagunit \omega m_p \mathds{1} + \left ( -\imagunit \omega m_p + \gamma_0 \right ) \left [ \mathds{1} + \left (1 + \delta + \frac{\delta^2}{3} \right ) \gamma_s \tens{R}
\right ]^{-1}.
\end{equation}
We observe that unlike with the other results, the drag coefficient depends on the mass of the particle $m_p$, which does not cancel out even if we expand to first order in
$\gamma_s \tens{R}$.  The drag coefficients from the modified point-particle framework are calculated from~\eqref{eq:DragCoefficientTensorFromReactionFieldTensor} using
the expressions for the reaction field tensor from~\citet[equations (3.5) and (3.16)]{felderhof2012PhysRevE}, which we have reproduced in equation~\eqref{eq:FelderhofReactionField}.

Figure~\ref{fig:ReflectionsVsPPA} compares the real and imaginary parts of drag coefficients for a no-slip sphere near a full-slip wall obtained from the method of reflections,
and from the modified point-particle approximation for the case of $\epsilon = a/h = 0.5$. The free-space drag coefficient $\gamma_0(\omega)$ has been subtracted in order to
clearly show the difference between the methods. The inset in sub-figure (c) shows a log-log plot of $\Real (\gamma_\| / \gamma_s)$, i.e. without subtraction of the free-space
drag coefficient, exemplifying the excellent agreement between the two methods even for the large value of $\epsilon$. \suppress{The relative discrepancy between these two methods,
calculated as $|\gamma^R - \gamma^P|/|\gamma^P|$ and expressed as a percentage, is plotted in figure~\ref{fig:ReflectionsVsPPAError} for different values of $\epsilon$. As
expected, the discrepancy becomes larger as the value of the small parameter $\epsilon$ increases, i.e. as the particle is moved closer to the wall, but is still quite small
even for $\epsilon = 0.5$ when the particle's centre is one diameter away from the wall.}

Figure~\ref{fig:PPAVsFelderhof} compares the real and imaginary parts of drag coefficients for a no-slip sphere near a full-slip wall obtained from Felderhof's point-particle
approximation, and from the modified point-particle framework, for the case of $\epsilon = 0.5$ and $\rho_p = 19 \rho_f$. If the liquid is water, this density corresponds
roughly to that of gold particles. As before, the free-space drag coefficient $\gamma_0(\omega)$ has been subtracted in order to clearly highlight the disagreement between the methods
at high frequencies. The inset in sub-figure (c) shows a log-log plot of $\Real (\gamma_\| / \gamma_s)$, i.e. without subtraction of the free-space drag coefficient,
showing that there is still visible disagreement between the two methods for large $\rho_p / \rho_f$. The relative error between these two approximations, calculated as
$|\gamma^F - \gamma^P|/|\gamma^P|$ and expressed as a percentage, is plotted in figure~\ref{fig:PPAVsFelderhofError} for different values\footnote{If the liquid is water, the values 2, 4 and 19 for $\rho_p/\rho_f$ roughly correspond to particles made of silica glass, Barium Titanate glass, and gold respectively. These are common choices in optical tweezers experiments.} of $\rho_p$. The error is zero when
$\rho_p = \rho_f$, and the errors become larger as $\rho_p$ deviates from $\rho_f$.

Figure~\ref{fig:AddedMassCheck} shows the high-frequency behaviour of the imaginary components of the drag coefficients from Felderhof's version, and from the modified version
of the point-particle approximation on a log-log scale. A line corresponding to the added mass contribution predicted from potential flow~\citep{milne1968theoretical} is
shown. The plots show the agreement of the modified point-particle approximation with the potential flow results at high frequencies.

\suppress{
\begin{figure}
\centering
\includegraphics[width=\textwidth]{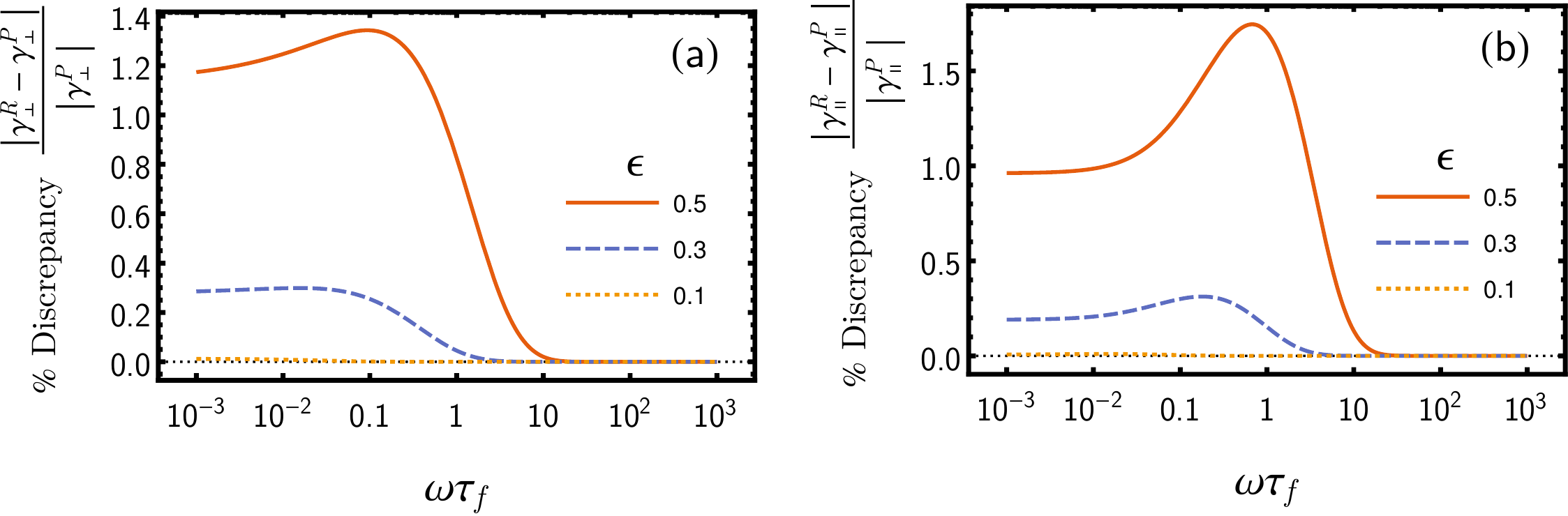}
\caption{\label{fig:ReflectionsVsPPAError}Semi-logarithmic plots of the percentage discrepancy between the drag coefficients obtained through the two approximation methods
  plotted in figure~\ref{fig:ReflectionsVsPPA}, calculated using $100\% \left | \gamma^R - \gamma^P \right | / \left | \gamma^P \right |$ for various values of $\epsilon$ (a) in the perpendicular
    direction, (b) in the parallel direction to the wall. As expected, the discrepancy becomes very small at small $\epsilon$, i.e. when the particle is further from the wall.}
\end{figure}
}

\begin{figure}
\centering
\includegraphics[width=\textwidth]{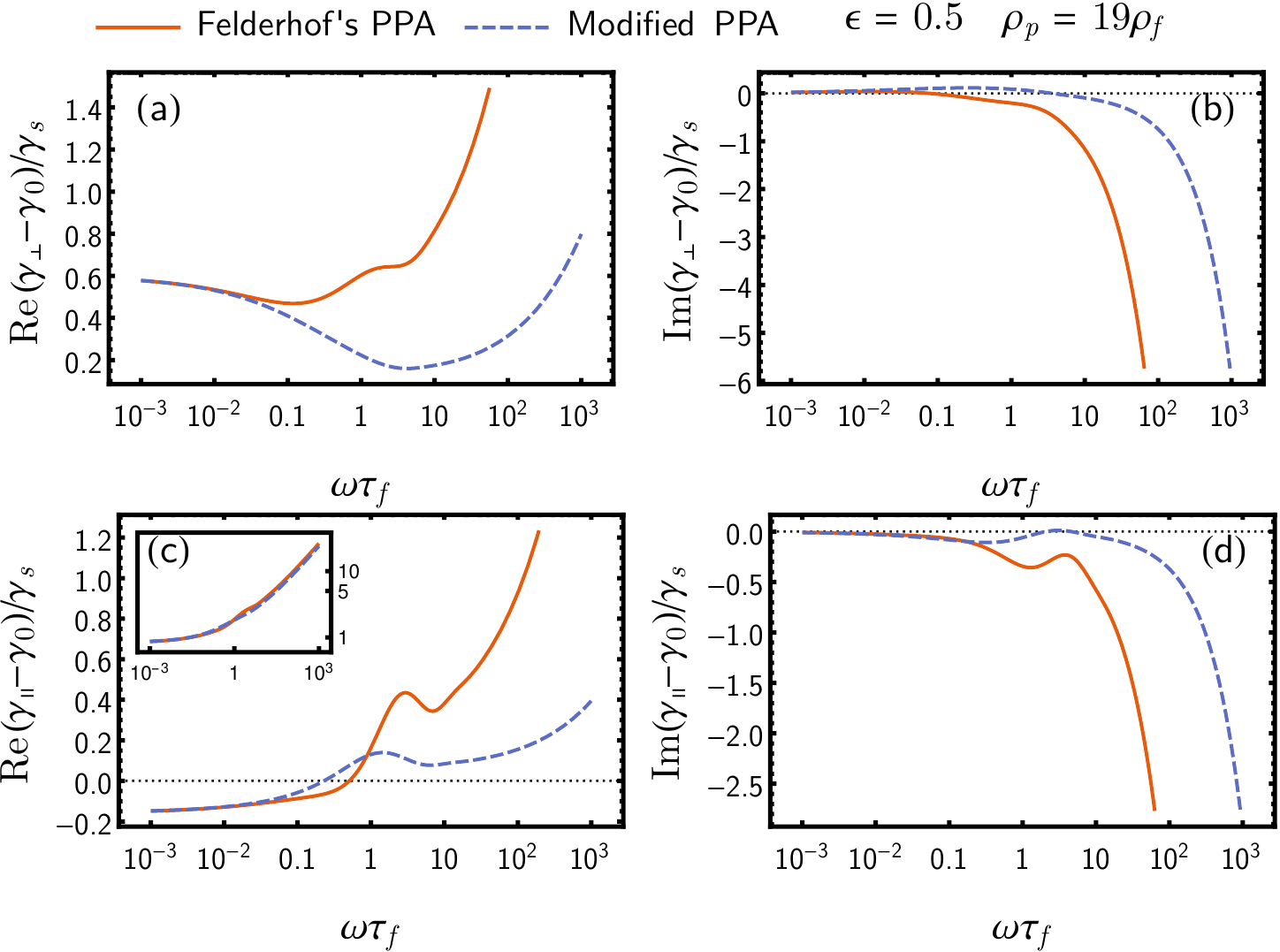}
\caption{\label{fig:PPAVsFelderhof}Semi-logarithmic plots comparing the results for the drag coefficient of a no-slip sphere near a full-slip plane wall ($\epsilon = 0.5$)
  from the point-particle approximation of~\citet{felderhof2012PhysRevE} (``Felderhof's PPA'') and the modified version of the point-particle approximation described in \S
  \ref{sec:point_particle_approximation} (``Modified PPA'') in the perpendicular direction (a) real part and (b) imaginary part, and in the parallel direction (c) real part
  and (d) imaginary part. In each case, the free-space drag coefficient $\gamma_0(\omega)$ has been subtracted in order to clearly highlight the differences, and the coefficients have
  been normalised by $\gamma_s$. Since the drag coefficient~\eqref{eq:DragCoefficientTensorFromAdmittanceUnmodified} from Felderhof's PPA depends on the density of the
  particle $\rho_p$, we set $\rho_p = 19 \rho_f$ (which is approximately the case for gold particles in water) to highlight the differences. The inset in (c) shows a log-log
  plot of the real parts of the normalised drag coefficients in the parallel direction without subtraction of $\gamma_0$.}
\end{figure}

\begin{figure}
\centering
\includegraphics[width=\textwidth]{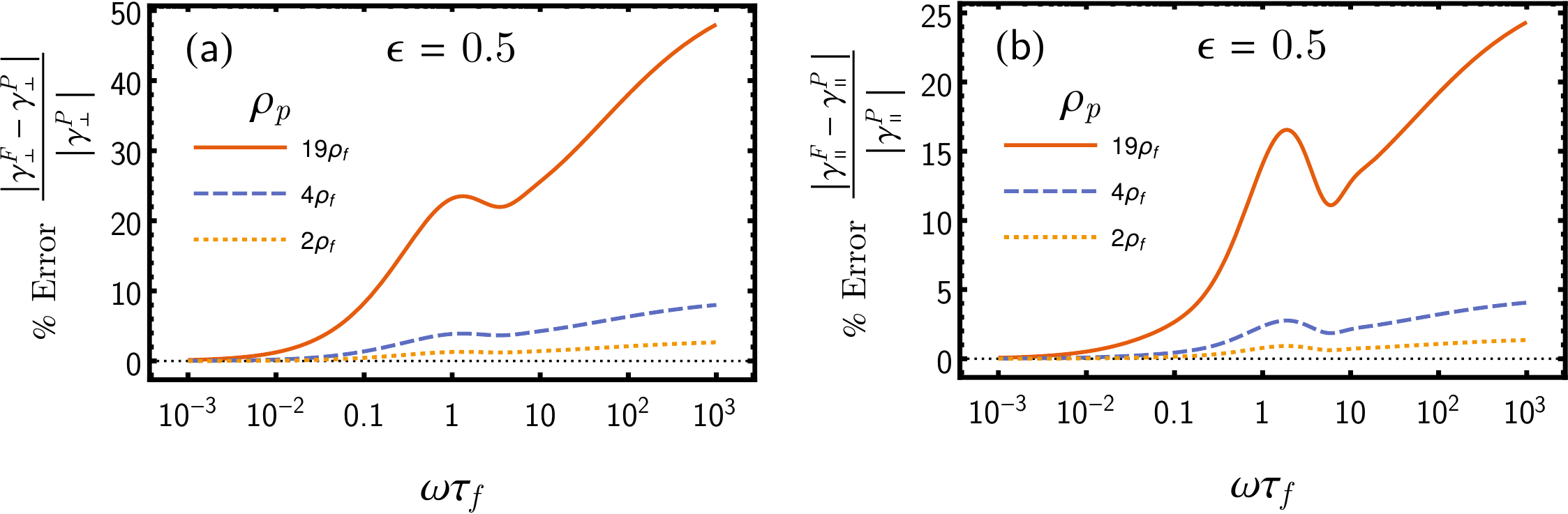}
\caption{\label{fig:PPAVsFelderhofError}Semi-logarithmic plots of the percentage error between the drag coefficients for a no-slip sphere near a full-slip plane wall obtained
  from Felderhof's PPA and the Modified PPA and calculated as $100\% \left | \gamma^F - \gamma^P \right | / \left | \gamma^P \right |$, for various values of particle density
  $\rho_p$ ($\epsilon = 0.5$) (a) in the perpendicular direction, (b) in the parallel direction to the wall.}
\end{figure}

\begin{figure}
\centering
\includegraphics[width=\textwidth]{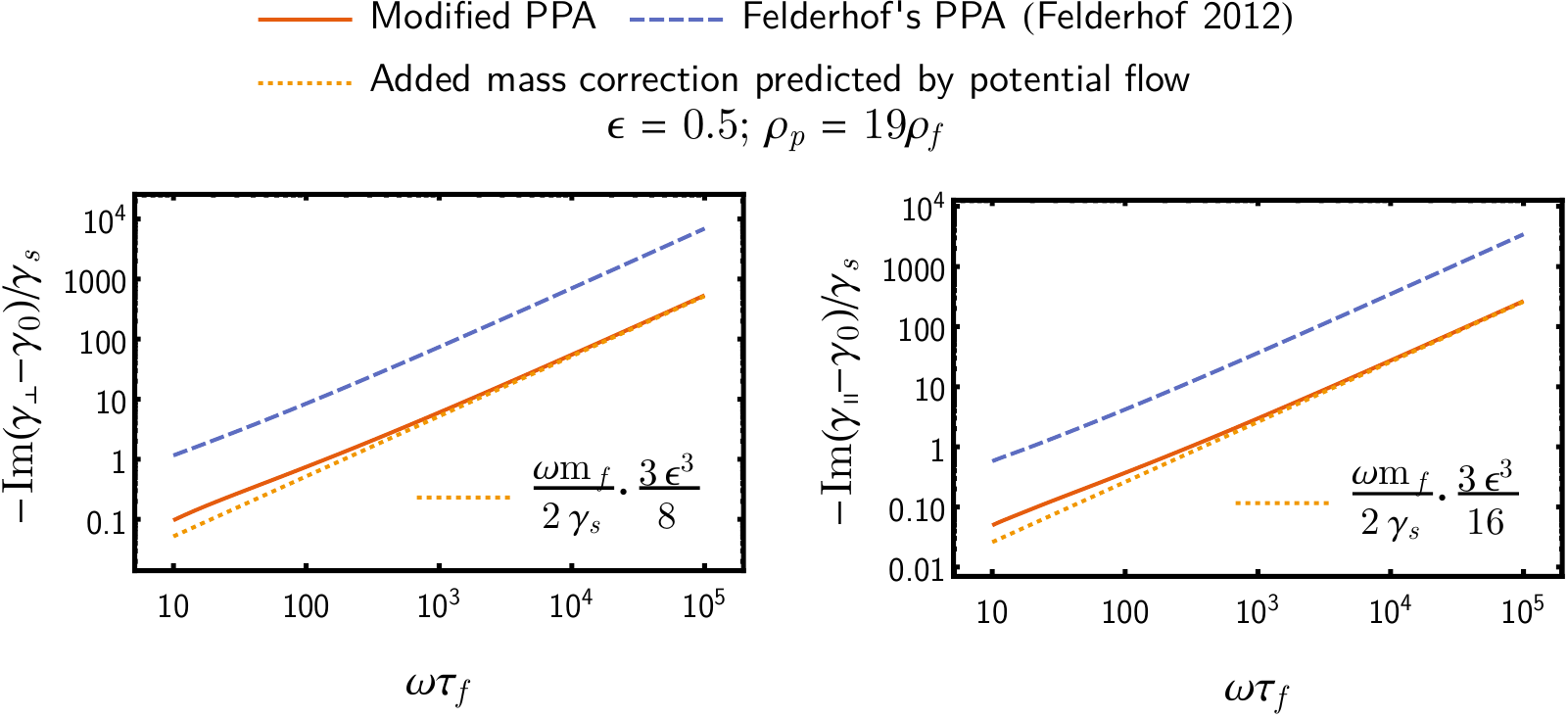}
\caption{\label{fig:AddedMassCheck}Logarithmic plots of $-\Imag(\gamma - \gamma_0)/\gamma_s$ for a no-slip sphere near a full-slip plane wall obtained from the modified
  (solid orange) and from Felderhof's original (dashed blue line) point-particle approximation against non-dimensionalized frequency $\omega \tau_f$. The region of slope
  $1$ of these lines represents the regime where the term corresponding to the increase in added mass due to the boundary is dominant. The dotted orange line plots the added
  mass correction from potential flow calculations. Values $\epsilon = 0.5$ and $\rho_p = 19 \rho_f$ are chosen to highlight the difference. It is observed that the modified
  point-particle approximation reproduces the results from potential flow at high frequencies.}
\end{figure}

\section{Application to Brownian Motion}
\label{sec:brownian_motion}

As discussed earlier, the short time-scale aspects of Brownian motion are relevant to fundamental science, microrheology, and to the calibration of instruments such as optical
tweezers, as thermal fluctuations play a significant role in these applications. In this section, we discuss the application of
these drag coefficient results to the problem of equilibrium Brownian motion of a spherical particle near a full-slip flat wall. We will also present the results for a no-slip
wall from the modified point-particle approximation, owing to its practical importance, and the discussion in~\citet[][\S II]{mo2015pre}. In both cases, we will analyse only
the translational motion, and ignore the rotational motion of the particle. This section includes a self-contained review of the theory, and a numerical analysis. More
detailed expositions of a general nature may be found in~\citet{li2013annalen} for example.

The long time-scale aspects of Brownian motion in a gas are well-modelled by employing the steady Stokes drag as a dissipation model. However, in a dense fluid, Brownian motion
is heavily influenced by the inertia of the fluid. Since the Reynolds number in many practical applications is very low ($\sim 10^{-4}$), it suffices for many purposes to
retain only the $\partial \vec{v} / \partial t$ and drop the advective term. Thus, unsteady Stokes friction provides a good model for the
dissipation~\citep{zwanzig1970hydrodynamicVACF,clercx1992brownian}. Of particular importance are the long-time power law tails of the velocity auto-correlation function, which
may be explained using the unsteady Stokes friction.

We note that to the approximation that $\dragtensor$ is diagonal, the equation of motion of the Brownian particle decouples into its Cartesian components, so we may treat the
motion perpendicular to and that parallel to a wall separately. This also holds true of the resulting predictions for statistical properties of the particle, such as power spectra
and auto-correlation functions. We also note that in our calculation of $\dragtensor$, we used the generalised Fax\'{e}n theorem of~\citet{mazur1974faxen}, which assumes that
the particle's boundary does not execute significant motion, whereby boundary conditions are applied on a stationary surface. This assumption would be
valid if the particle were confined by a tight potential~\citep{clercx1992brownian,franosch2009persistent}, which is fortunately indeed the case in many of the aforementioned practical
applications. Therefore, adding a harmonic restoring force to the equation of motion of the body bestows our model with theoretical consistency as well as enhances its
practical application. For simplicity, we shall assume that the tensor $\tens{K}$ of restoring force coefficients is diagonal in the basis suggested by the geometry of the sphere and wall.

Once the admittance~\eqref{eq:AdmittanceWithTrap} is known, the velocity auto-correlation function (VACF) $\tens{C}_v(t)$ of the Brownian particle may be calculated by inverting a
Green-Kubo relation to obtain~\citep[][eq. 19]{franosch2009persistent}
\begin{equation}
  \label{eq:GreenKubo}
\tens{C}_v(t) = \frac{2 k_B T}{\thepi} \int_0^\infty \mathrm{d}\omega\,\cos(\omega t)\,\Real[\admittance].
\end{equation}
The position auto-correlation function $\tens{C}_x(t)$ may be computed in a similar manner by using the mechanical susceptibility $\admittance / (-\imagunit \omega)$ in place
of the admittance in~\eqref{eq:GreenKubo}. Whereas the cosine transform in~\eqref{eq:GreenKubo} may be computed analytically for the case of a sphere in an unbounded medium,
one has to resort to numerical evaluation in most other cases\footnote{An algorithm for quadrature of oscillatory integrals such as a Filon-Trapezoid
  rule~\citep{tuck1967filon,franosch2009persistent} may be used.}.

The (two-sided) power spectral density of position ($\tens{S}_x$) and velocity ($\tens{S}_v$) fluctuations may also be computed through~\citep{franosch2009persistent,mo2015pre},
\begin{equation}
\begin{aligned}
\tens{S}_x(\omega) &= \frac{2 k_B T}{\omega^2} \Real[ \admittance ],\\
\tens{S}_v(\omega) &= 2 k_B T\, \Real[ \admittance ].
\end{aligned}
\end{equation}

Assuming $\dragtensor$ and $\tens{K}$ are diagonal, the mean-squared displacement (MSD) of the Brownian particle may be defined as
\begin{equation}
  \langle \Delta x_i^2(t) \rangle := \left < [ x_i(\tau + t) - x_i(\tau) ]^2 \right >
\end{equation}
for each component $x_i$ of the position $\vec{x}(t)$ of the particle. $\left < \; \right >$ denotes averaging over the ensemble of possible Brownian trajectories.
The mean-squared displacement may be related to the position auto-correlation function through
\begin{equation}
  \langle \Delta x_i^2(t) \rangle = 2 \left [ k_B T / K_{ii} - C_{x,ii}(t) \right ],
\end{equation}
where $K_{ii}$ and $C_{x,ii}$ denote the $i, i$ components of the diagonal tensors $\tens{K}$ and $\tens{C}_x$ respectively.

An alternative way to describe the Brownian motion of the particle is to use a stochastic equation of motion (often called a Generalised Langevin Equation) for the particle
\begin{equation}
\left[  -\imagunit \omega m_p \mathds{1} + \dragtensor - \tens{K} / (\imagunit \omega) \right ] \vec{u}_\omega = \vec{F}^\mathrm{th}_\omega,
\end{equation}
which we have written above in frequency domain. The Langevin force $\vec{F}^{\mathrm{th}}_\omega$ represents the effects of thermal fluctuations in the fluid, and is
typically modelled by a stationary stochastic process. In the Einstein-Ornstein-Uhlenbeck model of Brownian motion, which uses the steady Stokes drag, this stochastic process
is assumed to be white Gaussian noise. However, this choice is inconsistent with the fluctuation-dissipation theorem when the damping in the equation is
frequency-dependent. The theorem instead demands a coloured noise with a (two-sided) spectrum given by~\citep{balakrishnan1979fluctuation,franosch2011resonances},
\begin{equation}
\tens{S}_F(\omega) = 2 k_B T\, \Real[ \dragtensor ].
\end{equation}

It has been experimentally observed~\citep{mo2015pre} that the point-particle approximation using expression~\eqref{eq:DragCoefficientTensorFromReactionFieldTensor} performs
surprisingly well for large values of $\epsilon \approx 0.5$, i.e. when the particle is one diameter away from the wall. Motivated by this, when comparing predictions for
Brownian motion from the various theories (\S \ref{sec:brownian_motion}), we cast the expressions from the method of reflections in the (Pad\'{e}-like) form suggested by
equation~\eqref{eq:DragCoefficientTensorFromReactionFieldTensor} (see also \S\ref{sec:numerical_comparison}),
\begin{equation}
\label{eq:GammaReflectionsPade}
\frac{\gamma^R_{\perp,\|}}{\gamma_s} = \frac{\imagunit \omega m_f}{\gamma_s} + \frac{\left ( \tilde{\gamma}_0 / \gamma_s \right )^2}{\tilde{\gamma}_0 / \gamma_s - \Xi_{\perp,\|}(\epsilon, \delta)} + o[\Xi_{\perp,\|}],
\end{equation}
where the correction terms for the full-slip wall are given by,
\begin{equation}
\begin{aligned}
  \Xi_{\perp}(\epsilon, \delta) &= \frac{3 \epsilon}{8 \delta^2} \left [ \epsilon^2 \left ( 1 + \delta + \frac{\delta^2}{3} \right )^2 - \eulere^{2\delta(1-1/\epsilon)} \left (
      2 \epsilon \delta + \epsilon^2 \right ) \right ], \\
  \Xi_{\|}(\epsilon, \delta) &= \frac{3 \epsilon}{16 \delta^2} \left [ \epsilon^2 \left ( 1 + \delta + \frac{\delta^2}{3} \right )^2 - \eulere^{2\delta(1-1/\epsilon)} \left (
      4\delta^2 + 2 \epsilon \delta + \epsilon^2 \right ) \right ].
\end{aligned}
\end{equation}

We may now use the expressions for the admittance and the drag coefficients from the three theories and compare the predictions for the Brownian motion of a no-slip spherical
particle near a full-slip flat wall. We denote the components of the various diagonal tensors by $\perp$ and $\|$ just as we have done for the drag coefficient
tensor. To make connection with experiment, we will use exemplary parameters that are typical of optical tweezers
experiments~\citep[see][]{jeney2008boundary,mo2015pre,mo2015opex}. The same methods of numerical computation of the theoretical predictions described in~\S IV
of~\citet{mo2015pre} are employed here. It must be noted that it is not clear which of the two methods -- the modified point-particle approximation, or the method of
reflections -- performs better in practice, without higher order calculations or evidence from sophisticated experiments.

\begin{figure}
\centering
\includegraphics[width=\textwidth]{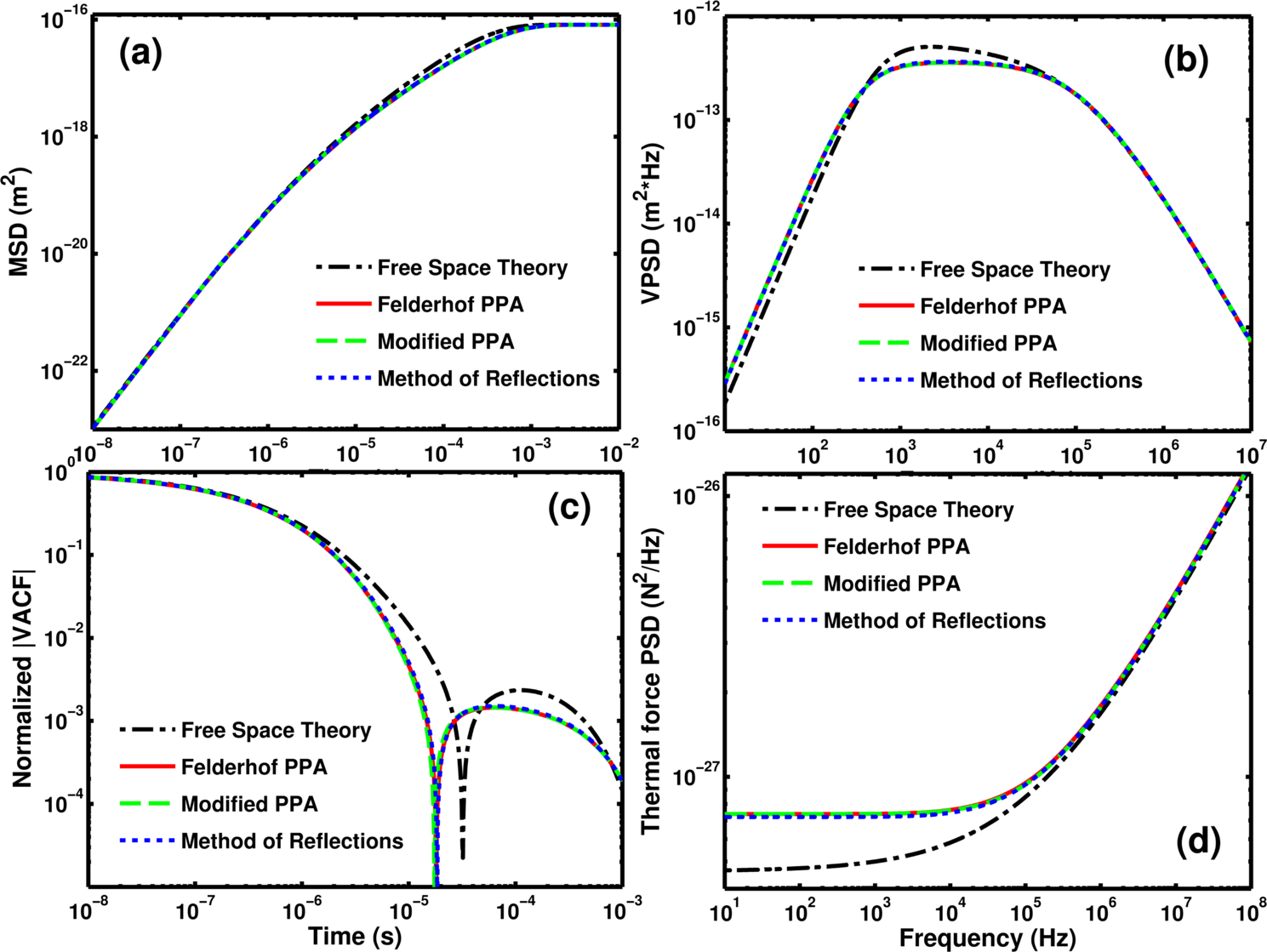}
\caption{\label{fig:PerpendicularBM}Logarithmic plots of predictions for the statistical properties of the Brownian motion of a silica glass sphere of $2a = 3$ $\mu$m diameter
  confined by a harmonic trap of stiffness $K = 100$ pN/$\mu$m at a distance of $h = 3$ $\mu$m from a full-slip plane wall in water, in the direction perpendicular to the wall. The solid red line shows the predictions using Felderhof's point-particle
  approximation without modification~\citep{felderhof2012PhysRevE}. The long-dashed green line shows predictions from the point-particle approximation with our modifications
  (\S~\ref{sec:point_particle_approximation}). The short-dashed blue line shows predictions from the method of reflections (\S~\ref{sec:reflection_method}). The black line of
  varying dash length shows the predictions for a similar particle in unbounded fluid, i.e. in the absence of a wall. Sub-figure (a) shows the mean-squared displacement (MSD)
  $\left <\Delta x_\perp^2(t) \right >$ as a function of time, (b) shows the one-sided power spectral density of velocity fluctuations $2 S^\perp_v$, (c) shows the absolute value of the velocity
  auto-correlation function $C^\perp_{v}(t)$, normalised by $C^\perp_{v}(0) = k_BT/m^\ast_\perp$, and (d) shows the one-sided power spectral density of the Langevin force $2 S^\perp_F$. The cusps in sub-figure (c) correspond to zero crossings, and are a result of the presence of the harmonic trap. The method of reflections and
  point-particle approximation agree very well for these parameters, despite the large value of $\epsilon = 0.5$.}
\end{figure}

Figure~\ref{fig:PerpendicularBM} compares predictions for the statistical properties of Brownian motion (temperature $T = 295$ K) of a harmonically confined (trap stiffness $K = 100$ pN/$\mu$m)
spherical Silica glass (density $\rho_p = 2.0 \rho_f$) particle (diameter $2a = 3$ $\mu$m) near ($\epsilon = a/h = 0.5$) a full-slip wall in water (density $\rho_f = 1000$ kg/m$^3$, viscosity $\eta = 10^{-3}$
$\mathrm{Pa} \cdot \mathrm{s}$) in the perpendicular direction to the wall from three theories for the drag coefficient -- Felderhof's point-particle approximation, the
modified point-particle approximation, and the method of reflections. Also shown for comparison are the predictions using the free-space drag coefficient
$\gamma_0(\omega)$. Sub-figure (a) shows the mean-square displacement (MSD), (b) shows the (one-sided) power spectral density of velocity fluctuations
$2 S^\perp_v$, (c) shows the velocity auto-correlation function $C^\perp_v(t) = \langle u_\perp(\tau) u_\perp(t + \tau) \rangle$, and (d) shows the (one-sided) power spectrum
of the Langevin force $2 S^\perp_F$.

\begin{figure}
\centering
\includegraphics[width=\textwidth]{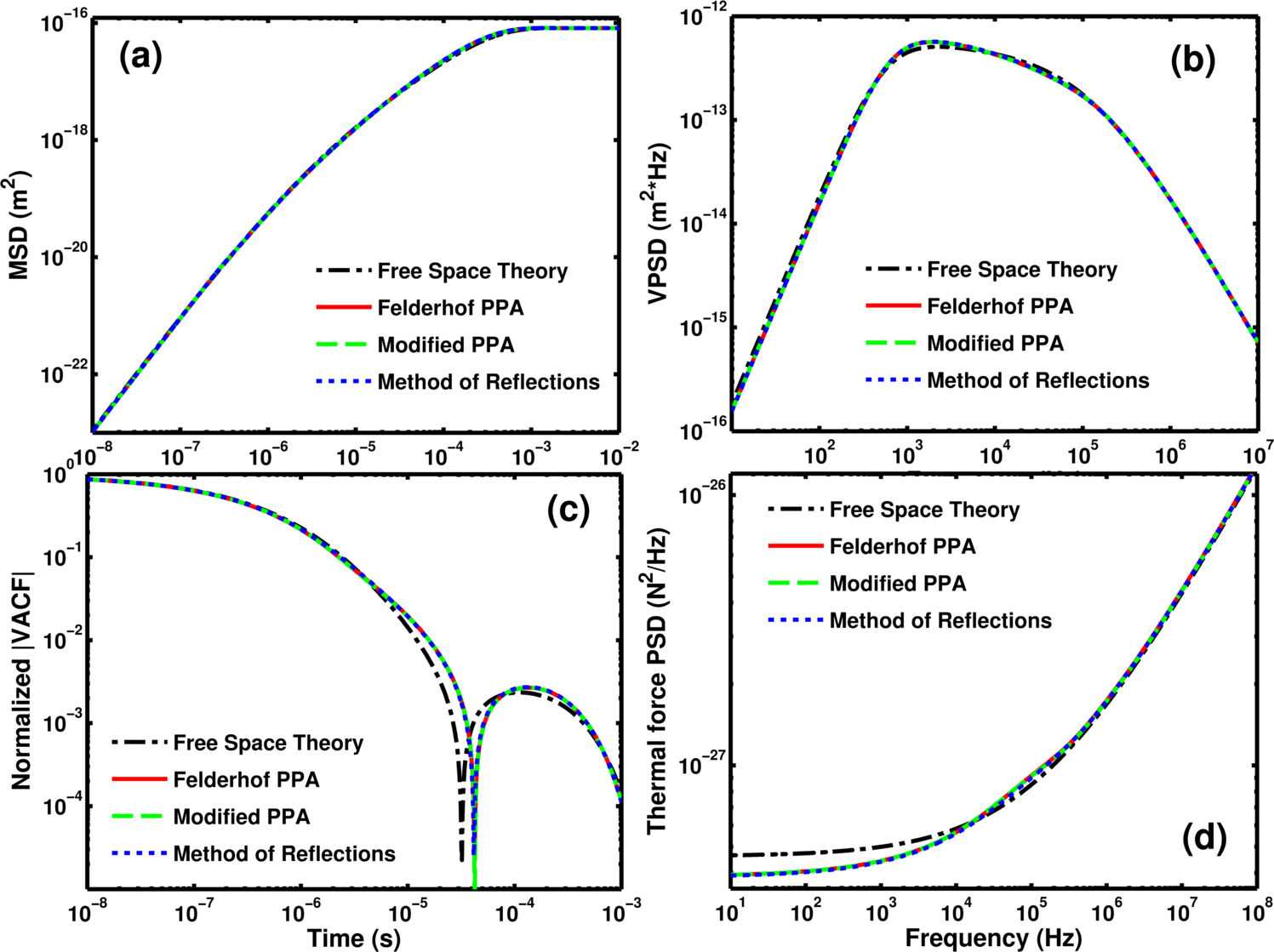}
\caption{\label{fig:ParallelBM}Logarithmic plots of predictions for the statistical properties of the Brownian motion of a silica glass sphere of $2a = 3$ $\mu$m diameter
  confined by a harmonic trap of stiffness $K = 100$ pN/$\mu$m at a distance of $h = 3$ $\mu$m from a full-slip plane wall in water, in the direction parallel to the wall. The solid red line shows the predictions using Felderhof's point-particle
  approximation without modification~\citep{felderhof2012PhysRevE}. The long-dashed green line shows predictions from the point-particle approximation with our modifications
  (\S~\ref{sec:point_particle_approximation}). The short-dashed blue line shows predictions from the method of reflections (\S~\ref{sec:reflection_method}). The black line of
  varying dash length shows the predictions for a similar particle in unbounded fluid, i.e. in the absence of a wall. Sub-figure (a) shows the mean-squared displacement (MSD)
  $\left <\Delta x_\|^2(t) \right >$ as a function of time, (b) shows the one-sided power spectral density of velocity fluctuations $2 S^\|_v$, (c) shows the absolute value of the velocity
  auto-correlation function $C^{\|}_{v}(t)$, normalised by $C^\|_{v}(0) = k_BT/m^\ast_\|$, and (d) shows the one-sided power spectral density of the Langevin force $2 S^\|_F$. The cusps in sub-figure (c) correspond to zero crossings, and are a result of the presence of the harmonic trap. The method of reflections and
  point-particle approximation agree very well for these parameters, despite the large value of $\epsilon = 0.5$.}
\end{figure}

Figure~\ref{fig:ParallelBM} compares predictions for the same statistical properties of Brownian motion for the same system, in the parallel direction to the full-slip wall
from the same three theories. As before, (a) shows the MSD, (b) shows the velocity PSD, (c) shows the VACF, and (d) shows the PSD of the Langevin force.

While it seems a formidable task to implement the method of reflections without approximation in the case of a no-slip wall, the calculation using the point-particle
approximation is tractable and has been accomplished by~\citet{felderhof2005JPhysChemB} (also see errata~\citet{felderhof2005erratum}). Felderhof's results for the reaction field tensor may be
employed in~\eqref{eq:DragCoefficientTensorFromReactionFieldTensor} to obtain predictions for the case of a no-slip wall that retain the benefits of our
modification~\citep[see][]{mo2015pre}.

\begin{figure}
\centering
\includegraphics[width=\textwidth]{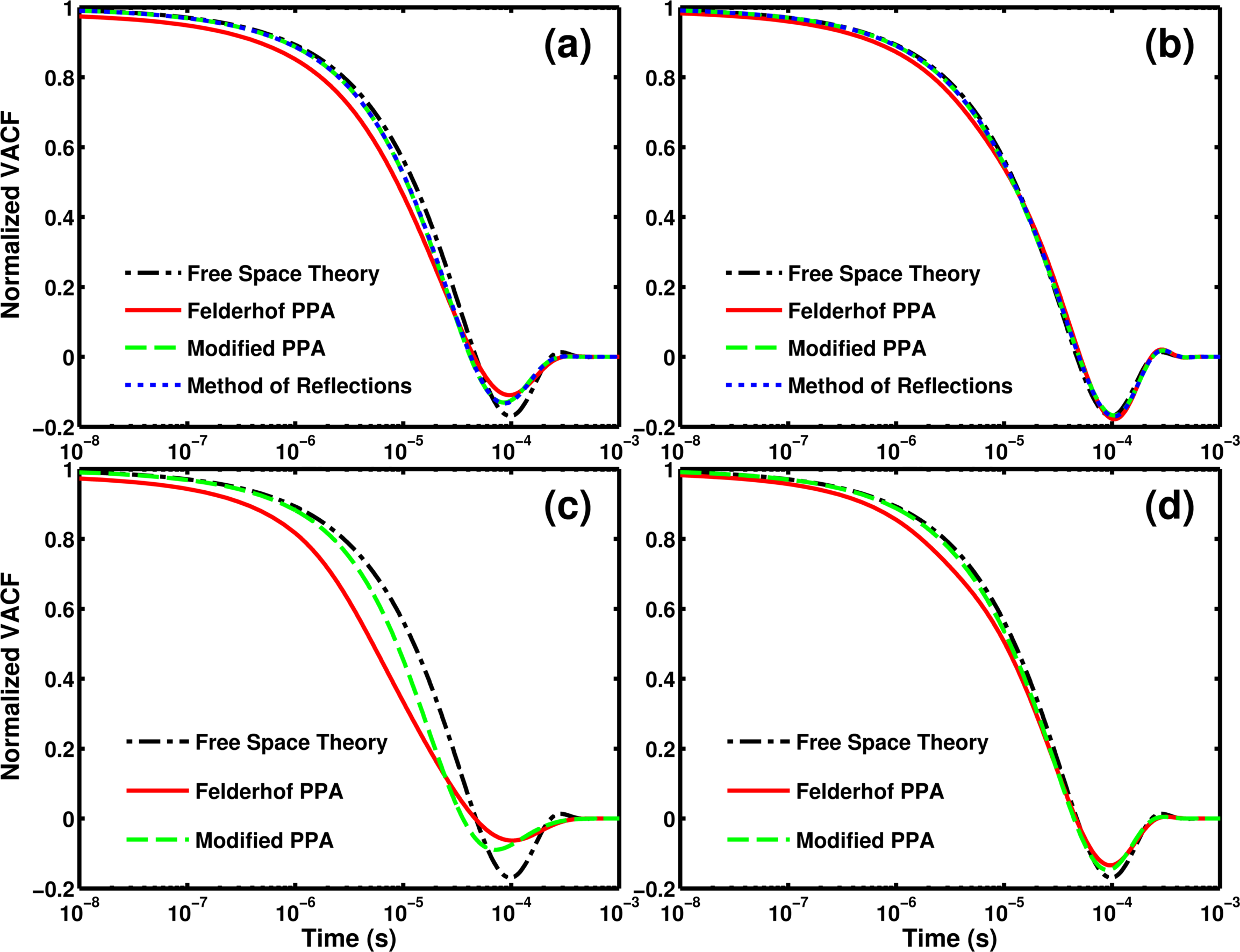}
\caption{\label{fig:VACFDiscrepancy}Semi-logarithmic plots of the predictions for the velocity auto-correlation function (VACF) of a $3$ $\mu$m diameter gold
  ($\rho_p = 19.3 \times 10^3$ kg/m$^3$) sphere in acetone ($\rho_f = 790$ $\mathrm{kg}/\mathrm{m}^3$), harmonically confined ($K = 200$ pN/$\mu$m) at a distance of $h = 3$ $\mu$m
  from a flat wall at a temperature $T = 295$ K. The solid red line shows the predictions using Felderhof's point-particle approximation without
  modification~\cite{felderhof2012PhysRevE,felderhof2005JPhysChemB,felderhof2005erratum}. The long-dashed green line shows predictions from the modified point-particle approximation
  (\S~\ref{sec:point_particle_approximation}). The short-dashed blue line in sub-figures (a) and (b) shows predictions from the method of reflections
  (\S~\ref{sec:reflection_method}). The black line of varying dash length shows the predictions for a similar particle in bulk fluid, i.e. in the absence of a
  wall. Sub-figures (a) and (b) show the results for a full-slip wall in the perpendicular and parallel directions respectively, and (c) and (d) show the same for a no-slip
  wall. The discrepancy between Felderhof's version of the point-particle approximation and our modified version indicates that our modifications would be important to systems of metallic particles in liquids.}
\end{figure}

Figure~\ref{fig:VACFDiscrepancy} shows the predictions for the VACF from the same three theories for the same temperature and geometry, but for gold
($\rho_p \approx 19.3 \times 10^{3}$ $\mathrm{kg}/\mathrm{m}^3$) particles in acetone ($\rho_f \approx 790$ $\mathrm{kg}/\mathrm{m}^3$) confined with a trap stiffness of
$K = 200$ pN/$\mu$m. The large ratio $\rho_p / \rho_f = 24.4$ is chosen to emphasise the dependence of the predictions from the theory
of~\citet{felderhof2005JPhysChemB,felderhof2005erratum,felderhof2012PhysRevE} on the particle density (the other two theories do not involve $\rho_p$). Sub-figures (a) and (b) show the results for
a full-slip wall in the perpendicular and parallel directions respectively. Sub-figures (c) and (d) show the results for a no-slip wall (the method of reflections is omitted
in this case).

It may thus be seen that in the regime of typical ($\rho_p \approx 2 \rho_f$) experiments using optical tweezers, it is difficult to distinguish between the theories,
explaining the agreement of previous experiments~\citep{jeney2008boundary} with theory of~\citet{felderhof2005JPhysChemB}. The experiment of~\citet{mo2015pre} uses the
modified point-particle approximation, but once again, does not constitute an experimental validation of any theory owing to the experimental uncertainty being larger than the
discrepancy between the theories. In the context of such systems, the modification would be of importance to high-precision measurements at sufficiently high frequencies,
possibly including lock-in measurements and precision measurements of statistical quantities with significant averaging.

Systems of gold and other metallic micro/nano-particles in liquids are common in experiments, not only in those involving optical tweezers~\citep[see
e.g.][]{svoboda1994,hajizadeh2010}, but also in other fields, given the wide array of applications of gold nano-particles~\citep{sardar2009}. Based on the results presented in
figure~\ref{fig:VACFDiscrepancy} for gold microspheres in acetone, we believe that our modification of the point-particle approximation would be very significant to such
systems.

\section{Discussion}
\label{sec:discussion}

In summary, our analysis shows that the modifications we introduced in \S \ref{sec:point_particle_approximation} are necessary to have theoretical consistency in the
predictions of Felderhof's point-particle approximation for the unsteady dynamics of a spherical particle in a liquid medium. As shown in \S \ref{sec:brownian_motion}, the
differences due to these modifications are too small to have been detected by previous experiments involving glass particles, but would be significant to experiments
involving metallic micro/nano-particles suspended in liquids.

Due to the presence of three length scales in the problem, the validity of the point-particle approximation in the context of unsteady Stokes flows needed further
scrutiny. Our formalisation of the point-particle approximation in \S \ref{sec:validity} shows that it is perturbatively consistent, and the comparison with the method of
reflections in \S \ref{sec:comparison_of_results} suggests that it may not capture all non-perturbative corrections. However, we have shown that these non-perturbative
corrections are small over the entire range of frequencies, explaining the excellent agreement with experiment.

Further work would use the point-particle approximation, with renewed confidence, in different geometries, potentially explaining the correction due to the curvature of a
cylindrical boundary observed in~\citet{mo2015pre}. Other avenues include developing similar frameworks to address rotational motion.

The authors wish to thank Prof. Mark G. Raizen for his extensive support, and Prof. Jean-Luc Thiffeault for many helpful discussions. A. S. acknowledges the hospitality of the
Geophysical Fluid Dynamics program, and support from a University of Texas at Austin Graduate Continuing Fellowship. J. M. acknowledges the support from the Sid W. Richardson
Foundation and the R. A. Welch Foundation Grant No. F-1258. P. J. M. acknowledges support from DOE Office of Fusion Energy Sciences under DE-FG02-04ER-54742, and from a
Forschungspreis from the Alexander von Humboldt Foundation. P. J. M. and A. S. would like to warmly acknowledge the hospitality of the Numerical Plasma Physics Division of the
Max Planck IPP, Garching, Germany.

\bibliographystyle{jfm}
\bibliography{refs}

\end{document}